\documentclass[a4paper,11pt]{article}
\pdfoutput=1 

\usepackage{jheppub} 

\usepackage[utf8]{inputenc}
\usepackage{graphicx}
\usepackage{dcolumn}
\usepackage{bm}

\renewcommand{\figurename}{{\bf Figure}}
\renewcommand{\thefigure}{{\bf\arabic{figure}}}
\renewcommand{\tablename}{{\bf Table}}
\renewcommand{\thetable}{{\bf\arabic{table}}}

\usepackage{natbib}
\usepackage{slashed}
\newcolumntype{x}[1]{>{\centering\arraybackslash\hspace{0pt}}p{#1}}
\usepackage{makecell}
\def\GeV{\,\text{GeV}}
\def\TeV{\,\text{TeV}}

\title{ \LARGE On the cosmological abundance of magnetic monopoles}

\author[a]{Chen Zhang,}
\author[a]{Shi-Hao Zhang,}
\author[a]{Bowen Fu,}
\author[a]{Jing-Fei Zhang}
\author[a,b,c,1]{and Xin Zhang}\note{Corresponding author.}

\affiliation[a]{Key Laboratory of Cosmology and Astrophysics (Liaoning) \& College of Sciences, Northeastern University, Shenyang 110819, China}
\affiliation[b]{Key Laboratory of Data Analytics and Optimization for Smart Industry (Ministry of Education), Northeastern University, Shenyang 110819, China}
\affiliation[c]{National Frontiers Science Center for Industrial Intelligence and Systems Optimization, Northeastern University, Shenyang 110819, China}

\emailAdd{zhangchen2@mail.neu.edu.cn}
\emailAdd{pumazhang200@163.com}
\emailAdd{fubowen@mail.neu.edu.cn}
\emailAdd{jfzhang@mail.neu.edu.cn}
\emailAdd{zhangxin@mail.neu.edu.cn}

\abstract{We demonstrate that Debye shielding cannot be employed to constrain the cosmological abundance of magnetic monopoles, contrary to
what is stated in the previous literature. Current model-independent bounds on the monopole abundance are then revisited for unit Dirac magnetic charge. We find that
the Andromeda Parker bound can be employed to set an upper limit on the monopole flux at the level of
$F_M\lesssim 5.3\times 10^{-19}\text{cm}^{-2}\text{s}^{-1}\text{sr}^{-1}$ for a monopole mass $10^{13}\GeV/c^2\lesssim m\lesssim 10^{16}\GeV/c^2$,
which is more stringent than the MACRO direct search limit by two orders of magnitude. This translates into stringent constraints on the monopole density
parameter $\Omega_M$ at the level of $10^{-7}\textendash 10^{-4}$ depending on the mass. For larger monopole masses the scenarios in which
magnetic monopoles account for all or the majority of dark matter are disfavored.}

\begin{document}

\maketitle

\section{Introduction}

Magnetic monopole is one of the most fascinating objects in theoretical physics~\cite{Preskill:1984gd,Stone:1984vom,Groom:1986ps,Giacomelli:2000de,Shnir:2005vvi,Balestra:2011ks,Rajantie:2012xh,Patrizii:2015uea,Patrizii:2019eud,Mavromatos:2020gwk}. It stems from the natural requirement
of extending the symmetry between electric and magnetic fields beyond the source-free Maxwell's equations. Its consistency
with quantum mechanics is demonstrated in 1931 by Dirac, which has led to an intriguing explanation for the charge quantization
observed in nature~\cite{Dirac:1931kp}. In 1974 't Hooft and Polyakov demonstrated that magnetic monopoles necessarily appear as topological defects
in particle theories of grand unification~\cite{tHooft:1974kcl,Polyakov:1974ek}, with the topological charge identified with the magnetic charge. It was soon realized
that the cosmological abundance of magnetic monopoles poses a serious problem for such scenarios~\cite{Zeldovich:1978wj,Preskill:1979zi,Vilenkin:2000jqa,Weinberg:2012pjx}, the solution of which gave
a strong boost to the birth of the inflation theory~\cite{Starobinsky:1980te,Guth:1980zm}. Because magnetic monopole is deeply connected
with the fundamental symmetries of nature at very high energy, the discovery of even a single magnetic monopole would have far-reaching
implications\footnote{We refer the reader to refs.~\cite{Bai:2021ewf,MoEDAL:2021vix,IceCube:2021eye,Iguro:2021xsu} for some recent interesting attempts at searching for magnetic monopoles.}. It is thus intriguing that the recent Amaterasu cosmic ray particle found by the Telescope Array Group~\cite{TelescopeArray:2023sbd} might be interpreted
as a magnetic monopole with extremely high energy~\cite{Cho:2023krz,Frampton:2024shp}.

Although it is possible that cosmic inflation has diluted away magnetic monopoles so that they end with an abundance
too small to have observable consequences, it is not necessarily the case~\cite{Preskill:1984gd,ParticleDataGroup:2022pth}. Relatively light magnetic monopoles might be produced
after inflation through various mechanisms~\cite{Kibble:1976sj,Zurek:1985qw,Murayama:2009nj,Das:2021wei,Kobayashi:2021des,Kobayashi:2022qpl,Maji:2022jzu,He:2022wwy,Lazarides:2024niy} so that their abundance is not affected by the dilution. Even if for superheavy
magnetic monopoles that are indeed affected by inflation, their final abundance still depends on its initial value and the actual amount of dilution and which
can be quite model-dependent and not well-constrained~\cite{Epele:2007ic,Senoguz:2015lba,Kephart:2017esj,Lazarides:2019xai,Chakrabortty:2020otp,Vento:2020vsq}. Alternative scenarios to inflation could also be envisioned~\cite{Brandenberger:2009jq}. Therefore
we choose to be open-minded and inquire what model-independent constraints can be obtained on the cosmological monopole abundance,
which is taken as a free parameter determined by some unknown production mechanism in the very early universe. Here by ``model-independent constraints''
we refer to constraints that can be derived from the electromagnetic and gravitational properties of the magnetic monopole only. Other properties
that depend on the model detail (such as catalysis of baryon number violation~\cite{Rubakov:1982fp,Callan:1982ah,Callan:1982au}) will not be assumed.

Recently it has been suggested in ref.~\cite{Medvedev:2017jdn} that Debye shielding in the magnetic monopole plasma leads to significant
constraints on the cosmological monopole abundance, especially for superheavy monopoles which are not well-constrained by
conventional means. Requiring the Debye length of the magnetic monopole plasma be larger than the scale of the largest coherent magnetic structure
so far observed, ref.~\cite{Medvedev:2017jdn} found that the cosmological density parameter of the magnetic monopole should satisfy
$\Omega_M\lesssim 3\times 10^{-4}$. This would certainly exclude the scenario in which magnetic monopoles account for all of the dark matter,
and would also be more stringent than the conventional Parker bound for relatively heavy nonrelativistic monopoles. It was expected that future
detection of magnetic fields of even larger coherent length would improve this bound by a few orders of magnitude.

In this work we shall however demonstrate that the Debye shielding bound suggested by ref.~\cite{Medvedev:2017jdn} cannot be used to obtain
meaningful bounds on the cosmological monopole abundance. The reason is that, as we shall see, the concept of Debye shielding simply does not
hold for magnetic monopole plasma in the presence of magnetic fields that are not curl-free. A similar issue in the context of solar physics
was discussed in Appendix B of ref.~\cite{Spicer:1984uui}, where the authors commented that ``Debye shielding will not shield out a transverse
electric field.'' In our analysis we note that Debye shielding is implicitly based on the requirement of thermal and mechanical equilibrium\footnote{{We discuss the subtle issue of thermal equilibrium of a magnetic monopole plasma at the beginning of Sec.~\ref{sec:anatf}.}} which leads to inconsistencies in the magnetic case with test electric currents.

For non-relativistic magnetic monopoles that do not catalyze baryon number violation, the existing bounds on their cosmological abundance
then become quite limited. We revisit the mass density constraint, the bound from direct search in cosmic rays and the conventional Parker
bounds. We find that the Andromeda Parker bound which was once proposed to constrain the abundance of magnetic black holes~\cite{Bai:2020spd}, can be employed
to constrain the abundance of ordinary magnetic monopoles effectively.

This paper is organized as follows. In Sec.~\ref{sec:problem} we demonstrate the problem with Debye shielding in the magnetic monopole plasma, which
implies that in an astrophysical context no meaningful bounds can be obtained from this effect. We then turn to current model-independent bounds
on cosmological monopole abundance in Sec.~\ref{sec:bounds}, disregarding the bound from Debye shielding. We present our discussion and conclusions in Sec.~\ref{sec:dnc}.

\section{Problem with Debye shielding in the magnetic monopole plasma}\label{sec:problem}

\subsection{Debye shielding in an electron-proton plasma}

We first give a concise review of Debye shielding in an ordinary (electron-proton) plasma, paying special attention to the assumptions
underlying the derivation. Following ref.~\cite{Thorne:2017mcp}, a fixed test charge $Q$ is introduced into a plasma of electrons and protons
at temperature $T$. It is expected that the plasma particles will redistribute in space until an equilibrium is reached. Suppose the equilibrium
state is characterized by a stationary proton number density distribution $n_p(\textbf{r})$ and a stationary electron number density distribution $n_e(\textbf{r})$,
and also an electrostatic potential $\Phi(\textbf{r})$. Taking the location of the test charge $Q$ to be the origin, the Poisson equation for $\Phi(\textbf{r})$
reads
\begin{align}
\nabla^2\Phi=-4\pi(n_p-n_e)e-4\pi Q\delta(\textbf{r}),
\label{eqn:poisson}
\end{align}
where $e>0$ denotes the proton charge and the equation is written in Gaussian units\footnote{In this work we adopt Gaussian units for electromagnetism and occasionally take $c=1$ to simplify equations.}. When the equilibrium at temperature $T$ is reached,
the number density distributions $n_p(\textbf{r})$ and $n_e(\textbf{r})$ should be controlled by the corresponding Boltzmann factor, that is
\begin{align}
n_p & =\bar{n}\exp[-e\Phi/(k_B T)],\\
n_e & =\bar{n}\exp[+e\Phi/(k_B T)].
\label{eqn:npne}
\end{align}
Here $k_B$ denotes the Boltzmann constant, and $\bar{n}$ denotes the proton (or electron) number density in the plasma at spatial infinity,
which is just the mean proton (or electron) number density averaged over a sufficiently large volume. We may now plug Eqs.~\eqref{eqn:npne}
into Eq.~\eqref{eqn:poisson} for a solution of $\Phi$, assuming spherical symmetry so that $\Phi$ is a function of the radial distance
$r$ only. Analytical solutions can be found by considering $e\Phi\ll k_B T$, so that
\begin{align}
\exp[\mp e\Phi/(k_B T)]\simeq 1\mp e\Phi/(k_B T).
\end{align}
The linearized Poisson equation then becomes
\begin{align}
\nabla^2\Phi=\frac{8\pi e^2\bar{n}}{k_B T}\Phi-4\pi Q\delta(\textbf{r}),
\end{align}
which has a spherically symmetric solution
\begin{align}
\Phi=\frac{Q}{r}e^{-\sqrt{2}r/\lambda_D},
\label{eqn:debyesolution}
\end{align}
with the Debye length $\lambda_D$ given by
\begin{align}
\lambda_D=\left(\frac{k_B T}{4\pi e^2 \bar{n}}\right)^{1/2}.
\end{align}
The solution Eq.~\eqref{eqn:debyesolution} has the desired property that (assuming $e\Phi\ll k_B T$ still holds)
\begin{align}
\Phi(r\rightarrow +\infty)=0,\quad \Phi(r\ll\lambda_D)=\frac{Q}{r},
\end{align}
while the appearance of the exponential suppression factor $e^{-\sqrt{2}r/\lambda_D}$ implies
that the electric field of the test charge $Q$ is screened beyond $r\sim\mathcal{O}(\lambda_D)$. In an intuitive picture
the test charge $Q$ repels plasma particles of the same sign while attracts plasma particles of the opposite sign, leading
to a net charged cloud of the opposite sign around $Q$, screening electric fields beyond a few Debye lengths.

\subsection{Potential issues for Debye shielding in the magnetic monopole plasma}

At first sight the above considerations carry over to the case of the magnetic monopole plasma (composed of monopoles and anti-monopoles) without difficulty, the only changes coming from
the magnetic coupling strength and the monopole mass. In fact this is true for test magnetic charges put into a quasi-neutral magnetic monopole plasma. Nevertheless, in a cosmological
or astrophysical setting, magnetic fields are likely to originate from electric currents rather than magnetic charges. For example, it is demonstrated in ref.~\cite{Parker:1987osc}
that the magnetic field of our own Galaxy is not associated with monopole oscillations. Therefore, to place bounds on monopole abundance, one should consider
test electric currents rather than test magnetic charges.

Now, both electric currents and magnetic charges are able to produce magnetic fields. However, they produce magnetic fields of distinct mathematical properties.
Electric currents are associated with magnetic fields that are divergencessless but not curl-free, referred to as ``transverse magnetic fields'' hereafter.
On the other hand, magnetic charges are associated with magnetic fields that are curl-free but not divergenceless, referred to as ``longitudinal magnetic fields''
hereafter. The key problem is therefore analyzing Debye shielding in the magnetic monopole plasma in the presence of transverse magnetic fields produced by
test electric currents. In this context, a number of issues arise compared to Debye shielding in an ordinary electron-proton plasma:
\begin{enumerate}
\item \textbf{Nonexistence of a scalar potential.} \quad The derivation of Debye shielding in an electron-proton plasma relies on the use of the electrostatic potential $\Phi$,
without which one cannot use a Boltzmann factor $\exp[\mp e\Phi/(k_B T)]$ to characterize the equilibrium proton and electron number density distribution. However, the
counterpart of $\Phi$, which should be a magnetic scalar potential, cannot be introduced since the magnetic fields from electric currents are not curl-free. One might
argue that a magnetic scalar potential can still be introduced in a simply connected region\footnote{A simply connected region is characterized by the fact that any closed loop
inside such a region can be contracted to a point in a continuous manner.} of space free of electric currents. However, this is certainly
unsatisfactory if one considers electric current densities that fill out a finite volume. This is also unsatisfactory when the region of space free of electric currents
is not simply-connected and thus needs to be covered by two or more simply-connected regions. A single-valued magnetic scalar potential cannot be assigned, prohibiting
the use of a Boltzmann factor.
\item \textbf{Screening of fields of distinct mathematical properties.} According to the discussion above, magnetic charges produce longitudinal magnetic fields while
electric currents produce transverse magnetic fields. Debye shielding then require transverse magnetic fields be screened by longitudinal magnetic fields, which is
quite counterintuitive.
\item \textbf{Lack of an intuitive physical picture.} \quad Debye shielding in an electron-proton plasma can be intuitively understood as originating from
attractive forces between opposite-sign charges and repulsive forces between same-sign charges. However, in the case of magnetic monopole plasma
with test electric currents, no such simple intuitive understanding is known.
\item \textbf{Two different interpretation of magnetic field neutralization.} \quad Debye shielding provide an alternative interpretation of magnetic field
neutralization in the presence of magnetic monopoles, which is different from the usual interpretation underlying the Parker bound~\cite{Parker:1970xv}. The conventional Parker bound
relies on the fact that magnetic monopoles drain energy from galactic magnetic fields. The complete physical picture should be understood as monopole plasma oscillation
with strong Landau damping~\cite{Turner:1982ag}. This is distinct from the Debye shielding interpretation, leading to different predictions of magnetic field
neutralization.
\end{enumerate}
Although the above four issues strongly question the validity of the concept of Debye shielding in a magnetic monopole plasma with test electric currents, they
do not lead to definite conclusions.

\subsection{Anatomy of the magnetic monopole plasma with a two-fluid model}
\label{sec:anatf}

In order to present a solid argument, let us show that the assumption that Debye shielding is achieved in a magnetic monopole plasma with test electric currents leads to inconsistencies, and thus cannot be used to obtain a meaningful constraint on the cosmological monopole abundance. We start by considering the magnetic monopole plasma using a two-fluid model, with one component fluid composed of magnetic monopoles, and the other component fluid composed of magnetic anti-monopoles. {This means that the two fluids can be characterized via the usual fluid equations and equations of state. Such a two-fluid treatment is routine practice in the
context of electron-ion plasmas in laboratory and space, while in the case of a magnetic monopole plasma one might
worry about the validity of such a fluid approximation. In ref.~\cite{Medvedev:2017jdn} the number of magnetic monopoles within a
Debye sphere is estimated to be much larger than unity in parameter space of phenomenological relevance\footnote{{Using notations and some benchmark values in Sec.~\ref{sec:bounds}, we may compute the number of magnetic monopoles within a Debye sphere as
\begin{equation}
N_D\simeq 2.3\times 10^{43}\times\left(\frac{mc^2}{10^{16}\GeV}\right)^{3/2}\left(\frac{v}{10^{-3}c}\right)^{7/2}\left(\frac{g}{g_D}\right)^{-3}
\left(\frac{F_M}{1.3\times 10^{-16}\,\text{cm}^{-2}\text{s}^{-1}\text{sr}^{-1}}\right)^{-1/2},\nonumber
\end{equation}
which is certainly much larger than unity in all parameter space of interest.}}, so collective effects
associated with the monopoles and anti-monopoles should be significant. On the other hand, the magnetic monopole plasma is
found to be highly collisionless~\cite{Medvedev:2017jdn}, since the collision frequency due to magnetic Coulomb scattering is very small.
Whether collisionless systems can be treated within a fluid approximation is a subtle issue. Zero collision frequency
implies an inifinite mean free path, leading to absurd results of divergent coefficients of viscosity and heat conduction.
There are known circumstances in which a fluid approximation is inadequate for capturing the essential physics of such collisionless
systems~\cite{Weinberg:2003ur,Flauger:2017ged}.}

{In the problem at hand the essence of the fluid approximation is the use of a temperature and an equation of state (based on an
isotropic pressure) to characterize the magnetic monopole plasma. Then the fluid equations can be simply obtained from integrating
the collisionless Boltzmann equation (and its first velocity moment) over the monopole velocity distribution. The equation of
state plays the role of a closure relation so that we do not need to consider higher moments of the collisionless Boltzmann equation~\cite{Thorne:2017mcp}. The notion of a temperature and the associated equation of state \textit{seems} to hinge on a state of
thermal equilibrium that can only be achieved through binary collisions. On the other hand, astrophysical plasmas are often
collisionless on scales of interest, and collisionless (astrophysical) plasmas are routinely observed or assumed to exhibit
collisional, fluid-like behaviors (e.g. `Langmuir's Paradox')~\cite{Bret:2015qia}. It is conceivable that a number of mechanisms
exist to achieve collisionless thermalization. There can be wave-particle scattering due to plasma instabilities.\footnote{{A similiar
phenomenon is the scattering of cosmic rays by Alfv\'{e}n waves~\cite{Thorne:2017mcp}.}} For self-gravitating systems phase mixing and violent relaxation can settle the system quickly into a stationary state~\cite{2008gady.book.....B}. Though out of thermal equilibrium in the strictest sense, in the context of dark matter physics it is possible~\cite{Hansen:2004dg,Ling:2009eh} that the dark matter velocity distribution can be described by Tsallis distributions derived from nonextensive statistical mechanics, which permits the notion of a temperature~\cite{2013SSRv..175..183L}.}

{As to the magnetic monopole plasma, although the effect of Coulomb collisions between monopoles is negligible, motion of magnetic
monopoles is subject to the influence of the environmental gravitational field and wave-particle interaction which are
yet to be studied. There are two possibilities regarding the outcome of these collisionless thermalization mechanisms.
The first possibility is that they are inadequate to bring the magnetic monopole plasma into a state of quasi-equilibrium\footnote{{In this work we use the word ``quasi-equilibrium" (or sometimes the phrase ``thermal equilibrium") in the sense that the distribution function is approximately Maxwellian, or something like Tsallis distributions that can be derived from nonextensive statistical mechanics, so that the notion of a temperature can be used. In case plasma particles obey Tsallis distributions, Debye shielding in an electron-ion plasma can still be discussed~\cite{2014JPlPh..80..341L}.}} that can be characterized by the notion of a temperature and the associated equation of state. Since the concept of Debye shielding hinges on such a state of quasi-equilibrium and the notion of a temperature, in this first possibility Debye shielding cannot be used to yield a meaningful constraint on the cosmological monopole abundance, so the plasma constraint obtained by ref.~\cite{Medvedev:2017jdn} does not apply. The second possibility is that the collisionless thermalization mechanisms
can bring the magnetic monopole plasma into a state of quasi-equilibrium that can be characterized by the notion of a temperature and the associated equation of state. In this work, we will however show that \emph{even if this second possibility is realized,
Debye shielding still cannot be used to yield a meaningful constraint on the cosmological monopole abundance.}
}

{Keeping the second possibility in mind, we therefore proceed with the two-fluid model, and write down the Euler equations for the magnetic monopole fluid and the anti-monopole fluid:\footnote{{Here we neglect terms representing the viscous force. Inclusion of viscous force terms does not alter the conclusion of the argument.}}}
\begin{align}
n_+ m\Bigg[\frac{\partial \textbf{u}_+}{\partial t}+(\textbf{u}_+\cdot\nabla)\textbf{u}_+\Bigg] & =-\nabla P_+ -n_+ m\nabla\Psi +n_+g(\textbf{B}-\textbf{u}_+\times\textbf{E}),\label{eqn:Euler1}\\
n_- m\Bigg[\frac{\partial \textbf{u}_-}{\partial t}+(\textbf{u}_-\cdot\nabla)\textbf{u}_-\Bigg] & =-\nabla P_- -n_- m\nabla\Psi -n_-g(\textbf{B}-\textbf{u}_-\times\textbf{E}).
\label{eqn:Euler2}
\end{align}
In the above two equations, $m$ is the mass of the magnetic monopole (which is the same for the magnetic anti-monopole), $g$ is the magnetic charge of
the monopole (so the anti-monopole carries magnetic charge $-g$). $n_+,\textbf{u}_+,P_+$ are the number density, velocity, and the pressure of the magnetic
monopole fluid, with the corresponding quantities for magnetic anti-monopole fluid denoted by a subscript $-$ instead of $+$. $\Psi$ is the gravitational potential which we
put in for generality. $\textbf{B},\textbf{E}$ are
the macroscopic magnetic and electric field strengths, respectively. There are also equations of continuity for the two fluids:
\begin{align}
\frac{\partial n_+}{\partial t}+\nabla\cdot(n_+\textbf{u}_+) & =0,\label{eqn:continuity1}\\
\frac{\partial n_-}{\partial t}+\nabla\cdot(n_-\textbf{u}_-) & =0.
\label{eqn:continuity2}
\end{align}
Furthermore, the macroscopic mganetic and electric fields should satisfy Maxwell's equations in the presence magnetic charge and currents:
\begin{align}
\nabla\cdot\textbf{E} & = 4\pi\rho_e, \label{eqn:Maxwell1}\\
\nabla\cdot\textbf{B} & = 4\pi\rho_m, \label{eqn:Maxwell2}\\
\nabla\times\textbf{E} & = -4\pi\textbf{J}_m-\frac{\partial\textbf{B}}{\partial t}, \label{eqn:Maxwell3}\\
\nabla\times\textbf{B} & = 4\pi\textbf{J}_e+\frac{\partial\textbf{E}}{\partial t}.
\label{eqn:Maxwell4}
\end{align}
Here $\rho_e,\textbf{J}_e$ are the electric charge and current densities respectively, while $\rho_m,\textbf{J}_m$ are the magnetic charge
and current densities respectively. Note that $\rho_m,\textbf{J}_m$ can be expressed in terms of $n_+,n_-,\textbf{u}_+,\textbf{u}_-$ as follows:
\begin{align}
\rho_m & =g(n_+ - n_-),\label{eqn:rhom}\\
\textbf{J}_m & =g(n_+\textbf{u}_+ - n_-\textbf{u}_-),
\label{eqn:Jm}
\end{align}
while $\textbf{J}_e$ is related to the macroscopic fields via Ohm's law:
\begin{align}
\textbf{J}_e = \sigma(\textbf{E}+\textbf{v}\times\textbf{B}),
\label{eqn:Ohm}
\end{align}
in which $\textbf{v}$ is the fluid velocity associated with the electric currents\footnote{$\textbf{v}$ should be thus related to $\textbf{J}_e$ via
an equation like Eq.~\eqref{eqn:Jm}, with an additional equation for the continuity of the electric charge. These details are not needed in the following
so we do not bother to present them explicitly.}, and $\sigma$ denotes the
electric conductivity which may be a function of the position and time. Finally, there should be equations of state which express
the pressure as a function of the number density and temperature:
\begin{align}
P_+ & = P_+(n_+,T),\label{eqn:eos1}\\
P_- & = P_-(n_-,T).
\label{eqn:eos2}
\end{align}
If the monopole and anti-monopole fluids can be viewed as dilute gases, these equations of state should be well-approximated
by the ideal gas law $P=n k_B T$. Here we allow for general equations of state of the Onnes type
\begin{align}
P_\pm(n_\pm,T)=\sum_{i=1}^{+\infty} c_{\pm i}(T) n_\pm^i.
\label{eqn:Onnes}
\end{align}

The key assumptions underlying the above system of equations include:
\begin{enumerate}
\item The magnetic monopole plasma can be viewed using a two-fluid model, with one fluid composed of magnetic monopoles
and the other fluid composed of magnetic anti-monopoles. The two fluids interact with each other only through macroscopic
electric and magnetic fields.
\item The fluid velocities are assumed to be non-relativistic.
\end{enumerate}
Similar assumptions are made in the usual treatment of electron-proton plasma in a two-fluid model~\cite{Thorne:2017mcp}. {Besides the subtlety discussed at the beginning of this subsection,} the first assumption implies that we are effectively working at scales much larger than the mean separation between magnetic monopoles
and the magnetic monopole plasma is not very dense so that the monopole annihilation can be neglected.

The system of equations from Eq.~\eqref{eqn:Euler1} to Eq.~\eqref{eqn:eos2} can then be solved given appropriate initial or boundary conditions.
Since we wish to discuss Debye shielding in analogy with the corresponding effect in an ordinary electron-proton plasma, let us consider
an equilibrium configuration characterized by
\begin{align}
\textbf{u}_+=\textbf{u}_-=\textbf{0},
\label{eqn:static1}
\end{align}
and
\begin{align}
\frac{\partial n_+}{\partial t} & =\frac{\partial n_-}{\partial t}=0,\label{eqn:static2}\\
\frac{\partial\textbf{B}}{\partial t} & =\textbf{0},\quad\frac{\partial\textbf{E}}{\partial t}=\textbf{0}.\label{eqn:static3}
\end{align}
Moreover, in an astrophysical setting we will make the following simplified assumption: the electric conductivity $\sigma$ is
very large while $\textbf{E}+\textbf{v}\times\textbf{B}\simeq\textbf{0}$ so as to maintain a given finite $\textbf{J}_e$.
Then the two Maxwell's equations associated with the magnetic field become
\begin{align}
\nabla\cdot\textbf{B} & = 4\pi g(n_+ - n_-),\label{eqn:BMaxwell1}\\
\nabla\times\textbf{B} & = 4\pi\textbf{J}_e,\label{eqn:BMaxwell2}
\end{align}
and the two Euler equations for the monopole and anti-monopole fluids become
\begin{align}
-\nabla P_+ -n_+ m\nabla\Psi +n_+g\textbf{B} & =\textbf{0},\label{eqn:EulerStatic1}\\
-\nabla P_- -n_- m\nabla\Psi -n_-g\textbf{B} & =\textbf{0}.\label{eqn:EulerStatic2}
\end{align}
Eqs.~\eqref{eqn:BMaxwell1} and~\eqref{eqn:BMaxwell2} simply imply that for $n_+ - n_-$ and $\textbf{J}_e$ that tend to zero sufficiently rapidly
at infinity (which is true for discussing Debye shielding), the magnetic field strength can be expressed as (according to the Helmholtz theorem)
\begin{align}
\textbf{B}(\textbf{r})=-\nabla\Bigg[\int\frac{g(n_+ - n_-)}{|\textbf{r}-\textbf{r}'|}d\tau'\Bigg]
+\nabla\times\Bigg[\int\frac{\textbf{J}_e}{|\textbf{r}-\textbf{r}'|}d\tau'\Bigg].
\label{eqn:Helmholtz}
\end{align}
On the other hand, the Euler equations~\eqref{eqn:EulerStatic1} and~\eqref{eqn:EulerStatic2} can be cast as
\begin{align}
g\textbf{B} & =\frac{\nabla P_+}{n_+}+m\nabla\Psi,\label{eqn:EulerStatic3}\\
g\textbf{B} & =-\frac{\nabla P_-}{n_-}-m\nabla\Psi.\label{eqn:EulerStatic4}
\end{align}
Now, for equations of state which are of the Onnes type (c.f. Eq.~\eqref{eqn:Onnes}), one may write
\begin{align}
\frac{\nabla P_+}{n_+} & =\nabla F_+,\quad F_+\equiv\int^{n_+}\frac{1}{n_+}\frac{\partial P_+}{\partial n_+}dn_+,\\
\frac{\nabla P_-}{n_-} & =\nabla F_-,\quad F_-\equiv\int^{n_-}\frac{1}{n_-}\frac{\partial P_-}{\partial n_-}dn_-.
\end{align}
Eqs.~\eqref{eqn:EulerStatic3} and ~\eqref{eqn:EulerStatic4} thus become
\begin{align}
g\textbf{B} & =\nabla(F_++m\Psi),\label{eqn:EulerStatic5}\\
g\textbf{B} & =-\nabla(F_-+m\Psi).\label{eqn:EulerStatic6}
\end{align}
If we take the curl of these two equations, the left-hand side becomes $4\pi g\textbf{J}_e$ while the right-hand side vanishes,
thus leading to inconsistencies. We therefore conclude that the concept of Debye shielding does not hold for the astrophysical
magnetic monopole plasma.

The above argument should hold obviously for volume current density $\textbf{J}_e$ which is nonzero in a macroscopic volume
across which the two-fluid description is valid for the magnetic monopole plasma. One might want to check the more singular case
when the test electric currents are given as a surface current density or simply a line current. In such cases the length scale
associated with the electric currents is smaller than the minimum scale at which the fluid equations are supposed to hold. Nevertheless,
one can still find solid arguments that invalidate the concept of Debye shielding. For example, consider a
steady line current running through a loop $L$. Now the region $\mathbb{R}^3\backslash L$ is not simply connected, prohibiting
the use of a single-valued magnetic scalar potential even if $\nabla\times\textbf{B}=\textbf{0}$ inside this region. However,
this problem can be circumvented if we choose an arbitrary finite surface $S$ bounded by $L$. Then the region $\mathbb{R}^3\backslash S$
is simply connected, which allows us to introduce a single-valued magnetic scalar potential $\Phi_m$. By the integral form
of Amp\`{e}re's law, $\Phi_m$ should exhibit a discontinuity across the surface $S$. This discontinuity should be proportional to
the electric current $I$ running through the loop $L$. The existence of such a discontinuity does not pose problem for defining
the magnetic field strength $\textbf{B}$, as the value of $\textbf{B}$ at the surface $S$ can be found by considering a different surface $S'$
bounded by $L$. However, it does pose a problem for discussing Debye shielding as we need to use the Boltzmann number density distributions
such as
\begin{align}
n_+ &=\bar{n}_m\exp[-g\Phi_m/(k_B T)],\\
n_- &=\bar{n}_m\exp[g\Phi_m/(k_B T)],
\end{align}
with $\bar{n}_m$ being the mean number density of the magnetic monopole (or aniti-monopole). These Boltzmann number density distributions are
solutions to the steady state Boltzmann equations, and the discontinuity in $\Phi_m$ across the surface $S$ results in discontinuities
in the number density distributions $n_\pm$ across $S$. Now the inconsistency is manifest: The surface $S$ can be chosen arbitrarily and thus
by choosing a different surface bounded by $L$ one ends up with a different continuity behavior for $n_\pm$. This is certainly not acceptable
since $n_\pm$ is physical. Moreover, discontinuity in $n_\pm$ across the surface $S$ in general also predicts a discontinuity in the pressure
$P_\pm$, leading to problem with mechanical equilibrium if we consider a small volume of fluid across the surface.

Inconsistencies also arise if we consider a surface current density flowing through a surface $S_1$ of finite area and bounded by
a loop $L_1$. The region $\mathbb{R}^3\backslash S_1$ is now simply connected, in which we can introduce a single-valued magnetic scalar potential
$\Phi_m$. However, if we construct an Amp\`{e}rian loop that crosses $S_1$ once at some point $P$, then according to the integral form of
Amp\`{e}re's law, $\Phi_m$ will in general exhibit a discontinuity across the surface $S_1$ at $P$. The amount of discontinuity depends on
the total electric current enclosed by the Amp\`{e}rian loop. According to the Boltzmann number density distributions, the discontinuity in
$\Phi_m$ results in discontinuity in $n_\pm$, which in turn results in discontinuity in the pressure $P_\pm$. The pressure discontinuity implies
that the mechanical equilibrium cannot be maintained if we consider a small volume of fluid across the surface $S_1$ around the point $P$. This argument can
be easily extended to the case that $S_1$ is a closed surface (such as the surface of a ball).

We therefore conclude that a stationary configuration of fluids and fields cannot be maintained at thermal and mechanical equilibrium under the assumptions of Eqs.~\eqref{eqn:static1}\textendash\textendash\linebreak~\eqref{eqn:static3}. These assumptions are obviously natural if we require magnetic Debye shielding be in parallel with
its electric counterpart. One assumption that one might wish to give up is the vanishing of the fluid velocities Eq.~\eqref{eqn:static1}. If we allow for
time-independent but spatially inhomogeneous fluid velocity fields $\textbf{u}_\pm$, we may get extra terms in Euler equations which could save their
compatibility with the Maxwell's equations. That is, Eqs.~\eqref{eqn:EulerStatic5} and~\eqref{eqn:EulerStatic6} become
\begin{align}
g\textbf{B} & =\nabla(F_++m\Psi)+m(\textbf{u}_+\cdot\nabla)\textbf{u}_+,\label{eqn:EulerStatic7}\\
g\textbf{B} & =-\nabla(F_-+m\Psi)-m(\textbf{u}_-\cdot\nabla)\textbf{u}_-.\label{eqn:EulerStatic8}
\end{align}
Taking the curl of these equations will not lead to inconsistencies. Furthermore, equations of continuity should be satisfied
\begin{align}
\nabla\cdot(n_+\textbf{u}_+) & =0,\label{eqn:continuity3}\\
\nabla\cdot(n_-\textbf{u}_-) & =0.
\label{eqn:continuity4}
\end{align}
Note that since $\textbf{B}$ should be viewed as given by Eq.~\eqref{eqn:Helmholtz}, Eqs.~\eqref{eqn:EulerStatic7}\textendash\textendash~\eqref{eqn:continuity4} then become a system of $8$ equations for $8$ functions $\textbf{u}_+,\textbf{u}_-,n_+,n_-$, once the test electric current $\textbf{J}_e$ is specified. Given the compatibility and independence of
these equations, we expect a unique solution for a given $\textbf{J}_e$. However, if we attempt to construct the magnetic current density $\textbf{J}_m$ according to
Eq.~\eqref{eqn:Jm}, in general we would expect a nonzero $\textbf{J}_m$. By the electric Amp\`{e}re's law Eq.~\eqref{eqn:Maxwell3} this generates an electric field
which decays in a power-law manner (rather than exponentially) at spatial infinity. Since a large-scale electric field is incompatible with current observations,
we conclude such a solution is not of practical interest and thus nonzero fluid velocities do not save the concept of Debye shielding in an astrophysical context.

Let us comment that if we introduce an ideal test magnetic dipole (rather than a test electric current loop enclosing a finite area) into the magnetic monopole plasma, the concept of Debye shielding can be made valid. Suppose the test magnetic dipole is located at some point $P_1$, then the region $\mathbb{R}^3\backslash P_1$ is simply connected, which allows the introduction
of a single-valued magnetic scalar potential. The number density distributions then follow a well-defined Boltzmann rule, with no ambiguities or unacceptable discontinuities.
For a test electric current loop enclosing a finite area the same argument does not go through.

\section{Revisiting current bounds on the cosmological monopole abundance}\label{sec:bounds}

In this section we revisit current bounds on the cosmological monopole abundance. We shall consider magnetic monopoles
with a Dirac charge $g_D=\frac{1}{2e}$ and mass $m\gtrsim 10^{11}\GeV$ so that they have not been accelerated
to relativistic velocities by magnetic fields in galaxies or galaxy clusters. Since it is demonstrated that
Debye shielding cannot yield a meaningful constraint, we shall collect and compare the following bounds:
\begin{enumerate}
\item Mass density constraint.
\item Direct search in cosmic rays.
\item Parker bounds.
\end{enumerate}
All these bounds have been discussed to various extent in previous literature~\cite{Burdin:2014xma,Patrizii:2015uea,
Patrizii:2019eud,Mavromatos:2020gwk,ParticleDataGroup:2022pth,Kobayashi:2023ryr}.\footnote{There exist other interesting bounds that constrain the monopole flux or simply the existence of
magnetic monopole in the spectrum of the theory, for example bounds from heavy-ion collision~\cite{Gould:2017zwi,MoEDAL:2021vix}, neutron star~\cite{Gould:2017zwi,Hook:2017vyc}, cosmic ray collision with atmosphere~\cite{Iguro:2021xsu}. These bounds however only constrain light magnetic monopoles with $m\lesssim 100\TeV$, much lighter than the mass range considered in this work.} The emphasis of the discussion
here is to examine whether magnetic monopoles may account for all (or the majority) of the dark matter, and the
assumptions and uncertainties involved in the analysis. Moreover, motivated by previous studies on magnetic black holes~\cite{Bai:2020spd,Ghosh:2020tdu,Diamond:2021scl},
we will improve the existing Parker bound by considering magnetic fields from the Andromeda galaxy.

Bounds on the cosmological monopole abundance can be represented in various manners. In many cases one considers
magnetic monopoles (including anti-monopoles) populated with the number density $n_M$ and an isotropic average velocity $v$ in some volume defined
in a reference frame $S$. Here the reference frame and the volume depend on the type of local measurement under consideration.
For example, for direct search in cosmic rays, the reference frame is fixed on the Earth and the volume is associated with the size of
the solar system, while for the Parker bound from the galactic magnetic field, the reference frame is fixed on the Milky way galaxy
and the volume is associated with the galactic magnetic field region in the Milky way. The local isotropic flux of magnetic monopoles is
defined as
\begin{align}
F_M=\frac{n_M v}{4\pi},
\label{eqn:fluxdef}
\end{align}
Eq.~\eqref{eqn:fluxdef} means that if one considers an area element $dS$ with its normal direction $\textbf{n}$, then during a short time $dt$ the number of
monopoles (and anti-monopoles) that cross $dS$ with velocity direction in a small solid angle $d\Omega$ around $\textbf{n}$ is given by
$F_M dtdSd\Omega$. Therefore the unit of $F_M$ is written as $\text{cm}^{-2}\text{s}^{-1}\text{sr}^{-1}$. To gain an intuitive feeling
of the flux, we may compute the mean separation $d_M=n_M^{-1/3}$ of magnetic monopoles as a function of $F_M$ and $v$ as
\begin{align}
d_M\simeq 288\,\text{km}\times\left(\frac{v}{10^{-3}c}\right)^{1/3}\left(\frac{F_M}{10^{-16}\text{cm}^{-2}\text{s}^{-1}\text{sr}^{-1}}\right)^{-1/3},
\end{align}
which implies monopoles saturating the flux upper limit of the MACRO experiment~\cite{MACRO:2002jdv} (see Eq.~\eqref{eqn:MACRO} below) and moving with a virial velocity $v\sim 10^{-3}c$
the mean separation is about $300\,\text{km}$.

Alternatively, the cosmological monopole abundance is embodied by the density parameter $\Omega_M$ of magnetic monopoles (including anti-monopoles), which is defined as
\begin{align}
\Omega_M=\frac{\bar{\rho}_M}{\rho_\text{crit}}.
\end{align}
Here $\bar{\rho}_M$ denotes the energy density of magnetic monopoles averaged on cosmological scales, and
\begin{align}
\rho_\text{crit}=1.05\times 10^{-5}h^2\GeV\,\text{cm}^{-3}
\end{align}
is the current critical density of the universe~\cite{ParticleDataGroup:2022pth}, with $h=0.674$ being the scaling factor for Hubble expansion rate~\cite{Planck:2018vyg}.

In order to discuss the conversion between $F_M$ and $\Omega_M$, we introduce $\gamma\equiv(1-v/c)^{-1/2}$ which is the Lorentz boost factor
of the magnetic monopoles measured in the reference frame $S$ associated with the local measurement under consideration, and $\rho_M=n_M mc^2\gamma$
which is the local energy density of the magnetic monopoles in the $S$ frame. The conversion between $F_M$ and $\Omega_M$ is then given by
\begin{align}
F_M & =\frac{\rho_\text{crit}\Delta v}{4\pi\gamma mc^2}\Omega_M, \label{eqn:conversion1}\\
\Omega_M & =\frac{4\pi\gamma mc^2}{\rho_\text{crit}\Delta v}F_M, \label{eqn:conversion2}
\end{align}
where the overdensity $\Delta$ is defined by
\begin{align}
\Delta\equiv\frac{\rho_M}{\bar{\rho}_M}.
\end{align}
Strictly speaking $\bar{\rho}_M$ is defined in the rest frame of the cosmic microwave background (CMB), while $\rho_M$ is defined in the $S$ frame. In practice the relative motion of $S$
with respect to CMB is always non-relativistic, and thus we ignore this small difference. Moreover, the use of the relation $\rho_M=n_M mc^2\gamma$
implies that the velocity in $\gamma\equiv(1-v/c)^{-1/2}$ does not coincide with the velocity that appears in Eq.~\eqref{eqn:fluxdef} due to
an extended velocity distribution. Again we ignore this small difference as we work in the non-relativistic regime. We will consider two benchmark
values of $\Delta$. The first is $\Delta=1$, applicable when magnetic monopoles are distributed uniformly in the universe. Such a uniform distribution
is possible only if magnetic monopoles are accelerated by cosmic or astrophysical magnetic fields so that they do not bind with galaxies or galaxy clusters.
The second is $\Delta=2.86\times 10^5$, which is the ratio between the local dark matter density $\rho_{\odot}=0.36\GeV\,\text{cm}^{-3}$~\cite{Sofue:2020rnl}
and the average dark matter density of the universe $\rho_{\text{DM}}=1.26\times 10^{-6}\GeV\,\text{cm}^{-3}$~\cite{ParticleDataGroup:2022pth}.
The use of $\Delta=2.86\times 10^5$ is based on the assumption that magnetic monopoles cluster in the same way as dark matter, and even with this assumption
in mind the value $\Delta=2.86\times 10^5$ is strictly speaking only applicable to direct searches in cosmic rays which indeed depend on the local monopole
density. For Parker bounds derived from astrophysical magnetic fields the relevant region is on a much larger scale and thus could be associated with
a different $\Delta$. For Parker bounds based on the magnetic field of the Milky Way we ignore this difference for simplicity. Based on the above considerations,
we will express $\Delta$ as a simple function of the monopole mass
\begin{align}
\Delta=
\begin{cases}
1, & m\leq 10^{18}\GeV/c^2, \\
2.86\times 10^5, & m>10^{18}\GeV/c^2.
\end{cases}
\label{eqn:Delta}
\end{align}
Here the boundary between clustered and unclustered monopoles is taken to be $m=10^{18}\GeV/c^2$, as suggested by ref.~\cite{Turner:1982ag,Kobayashi:2023ryr}
as an approximate estimate. In practical situations we shall expect a smooth transition from $\Delta\simeq 1$ at large monopole mass values to
$\Delta\simeq\mathcal{O}(10^5)$ to small monopole mass values, with the transition range spanning $2\sim 3$ orders of magnitude. Since a definite
prediction of the transition behavior is lacking, we opt for a simple step function as Eq.~\eqref{eqn:Delta}, keeping in mind the large uncertainties
in $\Delta$ associated with the transition range.

The conversion between $F_M$ and $\Omega_M$ critically depends on the velocity $v$ of the magnetic monopoles. For monopoles clustered with the galaxy
($m\gtrsim 10^{18}\GeV/c^2$), $v\simeq 10^{-3}c$ which is the virial velocity of the galaxy. Due to the presence of galactic and intracluster magnetic fields
lighter magnetic monopoles can be accelerated to higher (and even relativistic) velocities. It is estimated that magnetic monopoles with masses around
$10^{16}\textendash\textendash 10^{17}\GeV$ can be accelerated to $v\simeq 10^{-2}c$~\cite{Medvedev:2017jdn,ParticleDataGroup:2022pth} which is the escape velocity of large galaxy clusters~\cite{Medvedev:2017jdn}, while
magnetic monopoles with masses smaller than $10^{11}\textendash\textendash 10^{13}\GeV$ are expected to be in the relativistic regime~\cite{Wick:2000yc,Burdin:2014xma,Medvedev:2017jdn,ParticleDataGroup:2022pth}.
The dependence of the average monopole velocity on its mass is associated with its complicated acceleration history~\cite{Wick:2000yc,Perri:2023ncd} and thus cannot be predicted precisely.

To facilitate comparison between different constraints, let us express Eqs.~\eqref{eqn:conversion1} and~\eqref{eqn:conversion2} using various benchmark values (we work in the non-relativistic
regime here so that $\gamma\simeq 1$; for $\Delta$ we consider two benchmark values $\Delta=1$ and $\Delta=2.86\times 10^5$ so there are two equations each for $F_M$ and $\Omega_M$)
\begin{align}
F_M & = 3.07\times 10^{-18}\text{cm}^{-2}\text{s}^{-1}\text{sr}^{-1}\Delta\Big(\frac{mc^2}{10^{18}\GeV}\Big)^{-1}\Big(\frac{v}{10^{-3}c}\Big) \nonumber \\
& \times\Big(\frac{\rho_\text{crit}}{1.05\times 10^{-5}h^2\GeV}\Big)\Big(\frac{h}{0.674}\Big)^{-2}\Big(\frac{\Omega_M}{0.27}\Big),\label{eqn:FMc1} \\
F_M & = 8.78\times 10^{-13}\text{cm}^{-2}\text{s}^{-1}\text{sr}^{-1}\Big(\frac{\Delta}{2.86\times 10^5}\Big)\Big(\frac{mc^2}{10^{18}\GeV}\Big)^{-1}\Big(\frac{v}{10^{-3}c}\Big) \nonumber \\
& \times\Big(\frac{\rho_\text{crit}}{1.05\times 10^{-5}h^2\GeV}\Big)\Big(\frac{h}{0.674}\Big)^{-2}\Big(\frac{\Omega_M}{0.27}\Big),\label{eqn:FMc2}
\end{align}
\begin{align}
\Omega_M & = 11.4\Delta^{-1}\Big(\frac{mc^2}{10^{18}\GeV}\Big)\Big(\frac{v}{10^{-3}c}\Big)^{-1} \nonumber \\
& \times\Big(\frac{\rho_\text{crit}}{1.05\times 10^{-5}h^2\GeV}\Big)^{-1}\Big(\frac{h}{0.674}\Big)^{2}\Big(\frac{F_M}{1.3\times 10^{-16}\text{cm}^{-2}\text{s}^{-1}\text{sr}^{-1}}\Big), \label{eqn:OMc1}\\
\Omega_M & = 4.00\times 10^{-5}\Big(\frac{\Delta}{2.86\times 10^5}\Big)^{-1}\Big(\frac{mc^2}{10^{18}\GeV}\Big)\Big(\frac{v}{10^{-3}c}\Big)^{-1} \nonumber \\
& \times\Big(\frac{\rho_\text{crit}}{1.05\times 10^{-5}h^2\GeV}\Big)^{-1}\Big(\frac{h}{0.674}\Big)^{2}\Big(\frac{F_M}{1.3\times 10^{-16}\text{cm}^{-2}\text{s}^{-1}\text{sr}^{-1}}\Big).\label{eqn:OMc2}
\end{align}

We are then ready to revisit a variety of bounds on the cosmological abundance of non-relativistic magnetic monopoles.

\textbf{1. Mass density constraint.} This is simply the requirement that the density parameter of non-relativistic magnetic monopoles
should not exceed that of dark matter, i.e.~\cite{ParticleDataGroup:2022pth}
\begin{align}
\Omega_M\leq 0.27.
\end{align}
This can be converted into a constraint on $F_M$ using Eqs.~\eqref{eqn:FMc1} and~\eqref{eqn:FMc2}.

\textbf{2. Direct search in cosmic rays.} For non-relativistic magnetic monopoles with magnetic charge $g=g_D$, the most stringent bound from direct searches in
cosmic rays come from the MACRO experiment~\cite{MACRO:2002jdv}, which provides $90\%\text{C.L.}$ upper bounds on the local monopole flux.
Precise values of the bounds depend on the monopole velocity. For our purposes we adopt
\begin{align}
F_M\leq 1.3\times 10^{-16}\,\text{cm}^{-2}\text{s}^{-1}\text{sr}^{-1},
\label{eqn:MACRO}
\end{align}
which can be converted into a bound on $\Omega_M$ using Eqs.~\eqref{eqn:OMc1} and~\eqref{eqn:OMc2}.
The MACRO bound does not apply to the highly relativistic case with a Lorentz boost factor $\gamma\gtrsim 10$~\cite{MACRO:2002jdv}, while
for moderately or extremely relativistic cases there are bounds from other direct search experiments such as IceCube~\cite{IceCube:2021eye}
and Pierre Auger~\cite{PierreAuger:2016imq} which turn out to be more stringent.

\textbf{3. Parker bounds.} Parker bounds are based on the neutralization of cosmological or astrophysical magnetic fields~\cite{Beck:2011he,
annurev:/content/journals/10.1146/annurev-astro-091916-055221,Vachaspati:2020blt} by magnetic
monopoles. There are various versions of Parker bounds~\cite{Parker:1970xv,Turner:1982ag,Rephaeli:1982nv,Adams:1993fj,Lewis:1999zm,Kobayashi:2022qpl}, based on different assumptions on the evolution history of magnetic fields. The most reliable one is the original version with refinement by M. S. Turner et al.~\cite{Turner:1982ag} (referred to as TPB hereafter),
\begin{align}
\small
F_M\leq
\begin{cases}
10^{-15}\text{cm}^{-2}\text{s}^{-1}\text{sr}^{-1}\left(\frac{g}{g_D}\right)^{-1}\left(\frac{B}{3\times 10^{-6}\text{G}}\right)\left(\frac{R}{10^{23}\text{cm}}\right)^{1/2}
\left(\frac{l_c}{10^{21}\text{cm}}\right)^{-1/2}\left(\frac{t_{\text{reg}}}{10^{15}\text{s}}\right)^{-1},& m\leq \hat{m},\\
10^{-14}\text{cm}^{-2}\text{s}^{-1}\text{sr}^{-1}\left(\frac{g}{g_D}\right)^{-2}\left(\frac{mc^2}{10^{18}\GeV}\right)\left(\frac{v}{10^{-3}c}\right)^2
\left(\frac{l_c}{10^{21}\text{cm}}\right)^{-1}\left(\frac{t_{\text{reg}}}{10^{15}\text{s}}\right)^{-1},& m>\hat{m},
\end{cases}
\label{eqn:PBMW1}
\end{align}
where $B,R,l_c,t_{\text{reg}}$ is the averaged magnetic field strength, the scale of the magnetized region, the coherent length of the magnetic field,
and the regeneration time of the magnetic field of the Milky Way respectively, with their TPB benchmark values displayed in the denominators. The
threshold mass value $\hat{m}$ is given by
\begin{align}
\hat{m}=10^{17}\GeV/c^2\left(\frac{g}{g_D}\right)\left(\frac{v}{10^{-3}c}\right)^{-2}\left(\frac{B}{3\times 10^{-6}\text{G}}\right)\left(\frac{R}{10^{23}\text{cm}}\right)^{1/2}
\left(\frac{l_c}{10^{21}\text{cm}}\right)^{1/2}.
\label{eqn:PBMW2}
\end{align}

The Parker bound can be improved by considering astrophysical magnetic fields of larger coherence length and longer regeneration time. One promising
candidate is the magnetic field of the Andromeda galaxy, whose parameters can be estimated as~\cite{Han:1998he,Fletcher:2003ec,Arshakian:2008cx}\footnote{We note that the dark matter density
of Andromeda is similar to that of Milky Way~\cite{Tamm:2012hw}.}
\begin{align}
B\simeq 5\times 10^{-6}\text{G},\quad R\simeq l_c\simeq 10\,\text{kpc},\quad t_{\text{reg}}\simeq 10\,\text{Gyr}.\quad (\text{Andromeda})
\label{eqn:Andromeda}
\end{align}
Here for $B$ we consider only the regular magnetic field~\cite{Fletcher:2003ec}. The Parker bound from magnetic fields of the Andromeda galaxy
was employed by ref.~\cite{Bai:2020spd,Ghosh:2020tdu,Diamond:2021scl} to constrain the abundance of extremal magnetic black holes. It was then pointed out in
ref.~\cite{Kobayashi:2023ryr} that whether such magnetic black holes remain clustered with the Andromeda galaxy is uncertain. Here we use the
Andromeda version of the Parker bound to constrain the abundance of ordinary magnetic monopoles with $g=g_D$. The boundary between clustered
and unclustered monopoles in the case of the Andromeda galaxy is expected to be at $m\simeq 10^{19}\GeV/c^2$~\cite{Kobayashi:2023ryr}, again with a transition range
spanning $2\sim 3$ orders of magnitude. Therefore somewhat large uncertainty on the overdensity $\Delta$ is involved if one wants to convert the Andromeda Parker bound on the monopole flux
at large monopole mass $m\sim 10^{19}\GeV/c^2$ to a corresponding bound on the monopole density parameter. For much lighter magnetic monopoles
with $m\lesssim 10^{19}\GeV/c^2$ it is likely that they are unclustered with $\Delta\simeq 1$.

Using the benchmark values in Eq.~\eqref{eqn:Andromeda} we may express the Andromeda Parker bound as
\begin{align}
\footnotesize
F_M\leq
\begin{cases}
5.3\times 10^{-19}\text{cm}^{-2}\text{s}^{-1}\text{sr}^{-1}\left(\frac{g}{g_D}\right)^{-1}\left(\frac{B}{5\times 10^{-6}\text{G}}\right)\left(\frac{R}{10\,\text{kpc}}\right)^{1/2}
\left(\frac{l_c}{10\,\text{kpc}}\right)^{-1/2}\left(\frac{t_{\text{reg}}}{10\,\text{Gyr}}\right)^{-1},& m\leq \hat{m},\\
10^{-18}\text{cm}^{-2}\text{s}^{-1}\text{sr}^{-1}\left(\frac{g}{g_D}\right)^{-2}\left(\frac{mc^2}{10^{18}\GeV}\right)\left(\frac{v}{10^{-3}c}\right)^2
\left(\frac{l_c}{10\,\text{kpc}}\right)^{-1}\left(\frac{t_{\text{reg}}}{10\,\text{Gyr}}\right)^{-1},& m>\hat{m},
\end{cases}
\label{eqn:PBAndromeda1}
\end{align}
where the threshold mass $\hat{m}$ is given by
\begin{align}
\hat{m}=5.3\times 10^{17}\GeV/c^2\left(\frac{g}{g_D}\right)
\left(\frac{v}{10^{-3}c}\right)^{-2}\left(\frac{B}{5\times 10^{-6}\text{G}}\right)\left(\frac{R}{10\,\text{kpc}}\right)^{1/2}\left(\frac{l_c}{10\,\text{kpc}}\right)^{1/2}.
\label{eqn:PBAndromeda2}
\end{align}

Strictly speaking the process of extracting energy from galactic magnetic fields by magnetic monopoles should be viewed as belonging
to the first quarter cycle of a magnetic monopole plasma oscillation~\cite{Turner:1982ag}. If magnetic monopoles spend more than
the first quarter cycle in a coherent domain of the magnetic field, it might be possible that their kinetic energy is returned to
the magnetic field again~\cite{Long:2015cza}, invalidating the Parker bound obtained thereof. Such a situation can be avoided if
the monopole plasma oscillation is subject to strong Landau damping, which occurs for non-relativistic monopoles when
\begin{align}
\frac{\omega_M}{k}\lesssim v,
\label{eqn:LDcondition}
\end{align}
where $v$ is the thermal velocity of monopoles, $k$ is the minimum wavenumber associated with the magnetic field, and $\omega_M$
is the monopole plasma (angular) frequency
\begin{align}
\omega_M=\left(\frac{4\pi g^2 n_M}{m}\right)^{1/2}.
\end{align}
Eq.~\eqref{eqn:LDcondition} means that strong Landau damping occurs when the phase velocity of the monopole plasma oscillation
becomes comparable or less than the monopole thermal velocity~\cite{Parker:1987osc}. The minimum wavenumber $k$ is given by
\begin{align}
k=\frac{2\pi}{l_c},
\end{align}
where $l_c$ is the coherence length of the magnetic field. Using Eq.~\eqref{eqn:fluxdef} we may then express the condition for
strong Landau damping as
\begin{align}
l_c\lesssim\frac{v}{2g}\left(\frac{mv}{F_M}\right)^{1/2}.
\label{eqn:LD2}
\end{align}
Using benchmark values, Eq.~\eqref{eqn:LD2} can be expressed as\footnote{To obtain the numbers, we use $g_D\simeq 8.2\times 10^{-7}\GeV^{1/2}\,\text{cm}^{1/2}$ in Gaussian
units, which can be derived from the Dirac quantization condition $eg_D=\hbar c/2$ and $\alpha=\frac{e^2}{\hbar c}$.}
\begin{align}
l_c\lesssim 100\,\text{kpc}\left(\frac{g}{g_D}\right)^{-1}\left(\frac{v}{10^{-3}c}\right)^{3/2}\left(\frac{mc^2}{10^{18}\GeV}\right)^{1/2}
\left(\frac{F_M}{1.3\times 10^{-16}\,\text{cm}^{-2}\text{s}^{-1}\text{sr}^{-1}}\right)^{-1/2}.
\end{align}
Therefore for monopoles with $m\gtrsim 10^{16}\GeV/c^2$ and $v\simeq 10^{-3}c$ corresponding to the virial velocity, if the flux is at or below the MACRO direct search limit,
then since for the Andromeda galaxy $l_c\simeq 10\,\text{kpc}$, it is guaranteed that the monopole plasma oscillation is subject to strong Landau damping, so the Andromeda Parker bound is valid.
Lighter monopoles with $m\lesssim 10^{16}\GeV/c^2$ are expected to be accelerated by galactic and intracluster magnetic fields to higher velocities so as to achieve larger values
of $mv^3$. We therefore expect their oscillation is also subject to strong Landau damping. Furthermore we note that magnetic monopole plasma oscillation is incompatible
with the observed features of the magnetic fields in spiral galaxies~\cite{Parker:1987osc}. We therefore do not expect the oscillation effect to invalidate the Parker bound obtained thereof.

\begin{figure*}[ht]
\centering
	\includegraphics[width=0.8\linewidth]{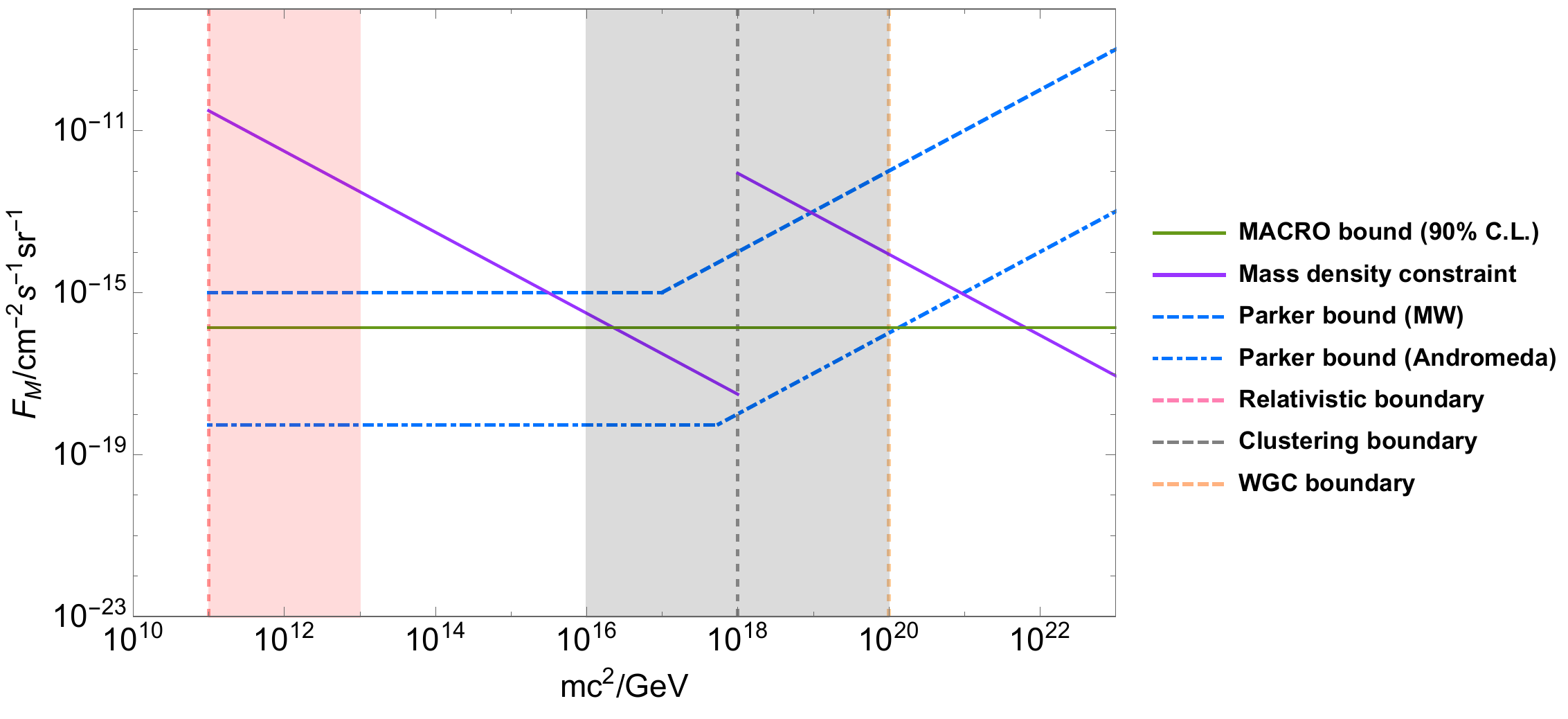}
	\caption{Bounds on the local flux $F_M$ of magnetic monopoles with $g=g_D$ as a function of monopole mass. The region below each line is excluded by the corresponding bound. The grey shaded area
is the estimated transition range from clustered to unclustered monopoles with large uncertainties in predicting the overdensity $\Delta$. The pink area is associated with
the uncertainty in predicting the relativistic boundary. See the text for details.}
	\label{fig:FM}
\end{figure*}

\begin{figure*}[ht]
\centering
	\includegraphics[width=0.8\linewidth]{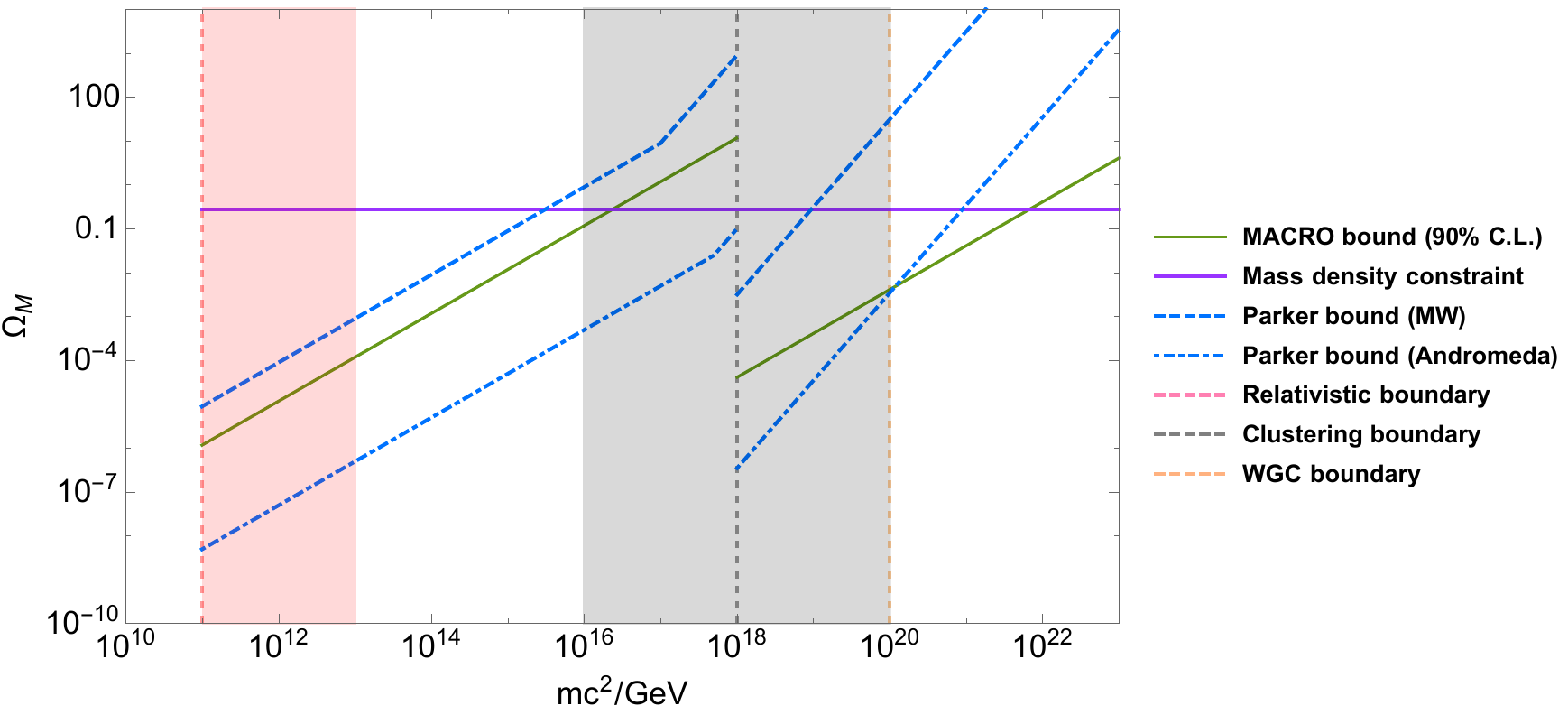}
	\caption{Same as Fig.~\ref{fig:FM}, but presented as bounds on the density parameter $\Omega_M$ of magnetic monopoles as a function of monopole mass.}
	\label{fig:OmegaM}
\end{figure*}

We summarize our findings in Figs.~\ref{fig:FM} and~\ref{fig:OmegaM}. The two figures contain the same information, but are displayed with
different vertical axes to facilitate interpretations. The conversion between two figures can be made by employing Eqs.~\eqref{eqn:FMc1}\textendash\textendash~\eqref{eqn:OMc2}.
In the figures, ``Parker bound (MW)'' is obtained from Eqs.~\eqref{eqn:PBMW1} and~\eqref{eqn:PBMW1} using benchmark values of the Milky Way shown in these equations, while
``Parker bound (Andromeda)'' is obtained from Eqs.~\eqref{eqn:PBAndromeda1} and~\eqref{eqn:PBAndromeda2} using benchmark values of the Andromeda galaxy. We do not display
bounds from Debye shielding proposed in ref.~\cite{Medvedev:2017jdn} since we have argued in Sec.~\ref{sec:problem} that the concept of Debye shielding is not valid for magnetic monopole
plasma in the presence of test electric currents. We focus on the case in which $g=g_D$ while results for other values of the magnetic charge can be inferred from simple
scalings.

Despite the uncertainties associated with the transition region for $\Delta$ and the relativistic boundary, important information can be extracted from
these figures. First consider the parameter region around $m\simeq 10^{20}\GeV/c^2$. This value of monopole mass is intriguing as it is the mass
predicted in Kaluza-Klein theory~\cite{Sorkin:1983ns,Gross:1983hb,Preskill:1984gd} and also the maximum mass allowed by the weak gravity conjecture (WGC)~\cite{Harlow:2022ich}.
Moreover, it is not constrained by the Milky Way Parker bound so at early times magnetic monopoles with masses around $m\simeq 10^{20}\GeV/c^2$ are conjectured
to make up all or the majority of dark matter~\cite{Preskill:1984gd}, despite the challenge in finding a convincing production mechanism.
The MACRO direct search experiment~\cite{MACRO:2002jdv} can however constrain such ultraheavy magnetic monopoles to an abundance one order of magnitude
smaller than that of dark matter. This direct search bound is subject to the astrophysical uncertainty in predicting the monopole density at scales comparable
to the solar system. This astrophysical uncertainty is important since the expected number of signal events in a detector module in the total collecting period
for a monopole flux saturating the MACRO bound is small (say $2\sim 3$). For direct detection of weakly interacting massive particle (WIMP) dark matter the corresponding
astrophysical uncertainty is not a big concern since simulations of dark matter fine-grained structures suggest a very smooth distribution~\cite{Schneider:2010jr,Vogelsberger:2010gd}.
For magnetic monopoles the situation is much less clear considering the plasma effects and astrophysical magnetism involved. The Andromeda Parker bound nevertheless
also disfavors the scenario in which such ultraheavy magnetic monopoles constitute all or the majority of dark matter.

Concerning the constraints on lighter monopoles with $10^{13}\GeV/c^2\lesssim m\lesssim 10^{18}\GeV/c^2$ (so that they remain non-relativistic), the MACRO bound
improves over the Milky Way Parker bound by roughly one order of magnitude, though the comparison is subject to astrophysical uncertainties to some extent. Moreover,
both the Milky Way Parker bound and the MACRO experiment do not constrain the mass range $10^{16}\GeV/c^2\lesssim m\lesssim 10^{18}\GeV/c^2$ well. The MACRO
bound does not seem to exclude monopoles within this mass range to account for all or the majority of dark matter. This situation is remedied by the
Andromeda Parker bound, which excludes such a possibility. Although this mass range is subject to uncertainties in $\Delta$, the conclusion remains unaffected
since the actual mass density constraint line in this mass range is supposed to be higher than what is shown in Fig.~\ref{fig:FM}. For lighter monopoles satisfying
$10^{13}\GeV/c^2\lesssim m\lesssim 10^{16}\GeV/c^2$ the uncertainties in $\Delta$ is much less relevant and we see that the Andromeda Parker bound improves over
its Milky Way counterpart by roughly three orders of magnitude, setting an upper limit on the monopole density parameter $\Omega_M$ to be about $10^{-7}\textendash 10^{-4}$
depending on the monopole mass.

\section{Conclusion}\label{sec:dnc}

Magnetic monopoles, if existing, would reflect fundamental symmetries of nature at very high energy, with profound implications
for both particle physics and cosmology. In this work we revisited a number of issues on their cosmological abundance. First
we demonstrated that Debye shielding cannot be employed to yield a meaningful constraint on the monopole abundance, contrary
to what is stated in the previous literature~\cite{Medvedev:2017jdn}. The concept of Debye shielding relies on a state of
thermal and mechanical equilibrium, which cannot be satisfied in the case of a magnetic monopole plasma in the presence of
test electric currents. This is shown explicitly by analyzing the corresponding fluid and Maxwell's equations. We then proceeded
to analyze three model-independent bounds on the cosmological monopole abundance, namely the mass density constraint, the direct
search constraint from MACRO experiment, and the Parker bounds. We have improved the Parker bound on magnetic monopoles
by considering magnetic fields of the Andromeda galaxy, which was previously used to constrain the abundance of magnetic black holes.
It is found that for non-relativistic monopoles with masses $10^{13}\GeV/c^2\lesssim m\lesssim 10^{16}\GeV/c^2$ and unit Dirac
magnetic charge, the Andromeda Parker bound is more stringent by the MACRO bound by two orders of magnitude, setting an upper limit
on monopole flux $F_M$ at the level of $5.3\times 10^{-19}\text{cm}^{-2}\text{s}^{-1}\text{sr}^{-1}$, or an equivalent upper limit
on the monopole density parameter $\Omega_M$ at the level of $10^{-7}\textendash 10^{-4}$ depending on the mass. For heavier monopoles
with masses $10^{16}\GeV/c^2\lesssim m\lesssim 10^{20}\GeV/c^2$ and unit Dirac magnetic charge the Andromeda Parker bound disfavors
scenarios in which such monopoles account for all or the majority of dark matter. Around $m\simeq 10^{20}\GeV/c^2$ which is the
largest mass allowed by the WGC, the Andromeda Parker bound is as competitive as the direct search limit from MACRO, albeit the comparison
between different bounds thereof hinges on important astrophysical uncertainties which deserve further investigation.

Magnetic monopoles can be produced in the early universe through symmetry-breaking phase transitions, which could be of first-order,
second-order, or a smooth crossover. For the monopole masses considered in this work, these phase transitions generically lead to
an unacceptably large monopole abundance, even if one takes into monopole annihilation into account~\cite{Preskill:1979zi}.
Inflation theory is the standard way to get rid of the unwanted monopoles, but if the inflationary energy scale is high,
lighter monopoles might still be produced, for example in partially-unified gauge theories~\cite{Pati:1974yy,Hartmann:2014fya,
DiLuzio:2020xgc,Dolan:2020doe,Cacciapaglia:2019dsq,Cacciapaglia:2020jvj,Lazarides:2021tua}. The subsequent evolution of their abundance
depends on a number of issues such as monopole annihilation, capture by primordial black holes~\cite{Stojkovic:2004hz,Zhang:2023tfv,
Zhang:2023zmb,Wang:2023qxj}, and interaction with other topological defects~\cite{Dvali:1997sa,Brush:2015vda}. Gravitational waves
might result from the symmetry-breaking phase transition if it is strongly first-order~\cite{Huang:2020bbe}, and solition
isocurvature perturbations~\cite{Lozanov:2023aez,Lozanov:2023knf}. Improving the model-independent bound on the cosmological monopole
abundance may thus have interesting implications for the early universe, which we leave for future study.

\begin{acknowledgments}
This work was supported by the National SKA Program of China (Grants Nos. \linebreak 2022SKA0110200 and 2022SKA0110203), the Joint Fund of Natural Science Foundation of Liaoning Province (Grant No. 2023-MSBA-067), the Fundamental Research Funds for the Central Universities (Grant No. N2405011), and the National 111 Project (Grant No. B16009).
\end{acknowledgments}

\bibliography{MA_v14}

\providecommand{\href}[2]{#2}\begingroup\raggedright\begin{thebibliography}{100}

\bibitem{Preskill:1984gd}
J.~Preskill, \emph{{MAGNETIC MONOPOLES}},
  \href{https://doi.org/10.1146/annurev.ns.34.120184.002333}{\emph{Ann. Rev.
  Nucl. Part. Sci.} {\bfseries 34} (1984) 461--530}.

\bibitem{Stone:1984vom}
J.~L. Stone, ed., \emph{{MONOPOLE '83. PROCEEDINGS, NATO ADVANCED RESEARCH
  WORKSHOP, ANN ARBOR, USA, OCTOBER 6-9, 1983}}, vol.~111, 1984.
\newblock 10.1007/978-1-4757-0375-7.

\bibitem{Groom:1986ps}
D.~E. Groom, \emph{{In Search of the Supermassive Magnetic Monopole}},
  \href{https://doi.org/10.1016/0370-1573(86)90037-2}{\emph{Phys. Rept.}
  {\bfseries 140} (1986) 323}.

\bibitem{Giacomelli:2000de}
G.~Giacomelli, M.~Giorgini, T.~Lari, M.~Ouchrif, L.~Patrizii, V.~Popa et~al.,
  \emph{{Magnetic monopole bibliography}},
  \href{https://arxiv.org/abs/hep-ex/0005041}{{\ttfamily hep-ex/0005041}}.

\bibitem{Shnir:2005vvi}
Y.~M. Shnir, \emph{{Magnetic Monopoles}}.
\newblock Text and Monographs in Physics. Springer, Berlin/Heidelberg, 2005,
  \href{https://doi.org/10.1007/3-540-29082-6}{10.1007/3-540-29082-6}.

\bibitem{Balestra:2011ks}
S.~Balestra, G.~Giacomelli, M.~Giorgini, L.~Patrizii, V.~Popa, Z.~Sahnoun
  et~al., \emph{{Magnetic Monopole Bibliography-II}},
  \href{https://arxiv.org/abs/1105.5587}{{\ttfamily 1105.5587}}.

\bibitem{Rajantie:2012xh}
A.~Rajantie, \emph{{Introduction to Magnetic Monopoles}},
  \href{https://doi.org/10.1080/00107514.2012.685693}{\emph{Contemp. Phys.}
  {\bfseries 53} (2012) 195--211},
  [\href{https://arxiv.org/abs/1204.3077}{{\ttfamily 1204.3077}}].

\bibitem{Patrizii:2015uea}
L.~Patrizii and M.~Spurio, \emph{{Status of Searches for Magnetic Monopoles}},
  \href{https://doi.org/10.1146/annurev-nucl-102014-022137}{\emph{Ann. Rev.
  Nucl. Part. Sci.} {\bfseries 65} (2015) 279--302},
  [\href{https://arxiv.org/abs/1510.07125}{{\ttfamily 1510.07125}}].

\bibitem{Patrizii:2019eud}
L.~Patrizii, Z.~Sahnoun and V.~Togo, \emph{{Searches for cosmic magnetic
  monopoles: past, present and future}},
  \href{https://doi.org/10.1098/rsta.2018.0328}{\emph{Phil. Trans. Roy. Soc.
  Lond. A} {\bfseries 377} (2019) 20180328}.

\bibitem{Mavromatos:2020gwk}
N.~E. Mavromatos and V.~A. Mitsou, \emph{{Magnetic monopoles revisited: Models
  and searches at colliders and in the Cosmos}},
  \href{https://doi.org/10.1142/S0217751X20300124}{\emph{Int. J. Mod. Phys. A}
  {\bfseries 35} (2020) 2030012},
  [\href{https://arxiv.org/abs/2005.05100}{{\ttfamily 2005.05100}}].

\bibitem{Dirac:1931kp}
P.~A.~M. Dirac, \emph{{Quantised singularities in the electromagnetic field,}},
  \href{https://doi.org/10.1098/rspa.1931.0130}{\emph{Proc. Roy. Soc. Lond. A}
  {\bfseries 133} (1931) 60--72}.

\bibitem{tHooft:1974kcl}
G.~'t~Hooft, \emph{{Magnetic Monopoles in Unified Gauge Theories}},
  \href{https://doi.org/10.1016/0550-3213(74)90486-6}{\emph{Nucl. Phys. B}
  {\bfseries 79} (1974) 276--284}.

\bibitem{Polyakov:1974ek}
A.~M. Polyakov, \emph{{Particle Spectrum in Quantum Field Theory}}, {\emph{JETP
  Lett.} {\bfseries 20} (1974) 194--195}.

\bibitem{Zeldovich:1978wj}
Y.~B. Zeldovich and M.~Y. Khlopov, \emph{{On the Concentration of Relic
  Magnetic Monopoles in the Universe}},
  \href{https://doi.org/10.1016/0370-2693(78)90232-0}{\emph{Phys. Lett. B}
  {\bfseries 79} (1978) 239--241}.

\bibitem{Preskill:1979zi}
J.~Preskill, \emph{{Cosmological Production of Superheavy Magnetic Monopoles}},
  \href{https://doi.org/10.1103/PhysRevLett.43.1365}{\emph{Phys. Rev. Lett.}
  {\bfseries 43} (1979) 1365}.

\bibitem{Vilenkin:2000jqa}
A.~Vilenkin and E.~P.~S. Shellard, \emph{{Cosmic Strings and Other Topological
  Defects}}.
\newblock Cambridge University Press, 7, 2000.

\bibitem{Weinberg:2012pjx}
E.~J. Weinberg, \emph{{Classical solutions in quantum field theory}: {Solitons
  and Instantons in High Energy Physics}}.
\newblock Cambridge Monographs on Mathematical Physics. Cambridge University
  Press, 9, 2012,
  \href{https://doi.org/10.1017/CBO9781139017787}{10.1017/CBO9781139017787}.

\bibitem{Starobinsky:1980te}
A.~A. Starobinsky, \emph{{A New Type of Isotropic Cosmological Models Without
  Singularity}},
  \href{https://doi.org/10.1016/0370-2693(80)90670-X}{\emph{Phys. Lett. B}
  {\bfseries 91} (1980) 99--102}.

\bibitem{Guth:1980zm}
A.~H. Guth, \emph{{The Inflationary Universe: A Possible Solution to the
  Horizon and Flatness Problems}},
  \href{https://doi.org/10.1103/PhysRevD.23.347}{\emph{Phys. Rev. D} {\bfseries
  23} (1981) 347--356}.

\bibitem{Bai:2021ewf}
Y.~Bai, S.~Lu and N.~Orlofsky, \emph{{Searching for Magnetic Monopoles with the
  Earth\textquoteright{}s Magnetic Field}},
  \href{https://doi.org/10.1103/PhysRevLett.127.101801}{\emph{Phys. Rev. Lett.}
  {\bfseries 127} (2021) 101801},
  [\href{https://arxiv.org/abs/2103.06286}{{\ttfamily 2103.06286}}].

\bibitem{MoEDAL:2021vix}
{\scshape MoEDAL} collaboration, B.~Acharya et~al., \emph{{Search for magnetic
  monopoles produced via the Schwinger mechanism}},
  \href{https://doi.org/10.1038/s41586-021-04298-1}{\emph{Nature} {\bfseries
  602} (2022) 63--67}, [\href{https://arxiv.org/abs/2106.11933}{{\ttfamily
  2106.11933}}].

\bibitem{IceCube:2021eye}
{\scshape IceCube} collaboration, R.~Abbasi et~al., \emph{{Search for
  Relativistic Magnetic Monopoles with Eight Years of IceCube Data}},
  \href{https://doi.org/10.1103/PhysRevLett.128.051101}{\emph{Phys. Rev. Lett.}
  {\bfseries 128} (2022) 051101},
  [\href{https://arxiv.org/abs/2109.13719}{{\ttfamily 2109.13719}}].

\bibitem{Iguro:2021xsu}
S.~Iguro, R.~Plestid and V.~Takhistov, \emph{{Monopoles from an Atmospheric
  Fixed Target Experiment}},
  \href{https://doi.org/10.1103/PhysRevLett.128.201101}{\emph{Phys. Rev. Lett.}
  {\bfseries 128} (2022) 201101},
  [\href{https://arxiv.org/abs/2111.12091}{{\ttfamily 2111.12091}}].

\bibitem{TelescopeArray:2023sbd}
{\scshape Telescope Array} collaboration, R.~U. Abbasi et~al., \emph{{An
  extremely energetic cosmic ray observed by a surface detector array}},
  \href{https://doi.org/10.1126/science.abo5095}{\emph{Science} {\bfseries 382}
  (2023) 903--907}, [\href{https://arxiv.org/abs/2311.14231}{{\ttfamily
  2311.14231}}].

\bibitem{Cho:2023krz}
Y.~M. Cho and F.~H. Cho, \emph{{Has Telescope Array Discovered Electroweak
  Monopole?}},  \href{https://arxiv.org/abs/2312.08115}{{\ttfamily
  2312.08115}}.

\bibitem{Frampton:2024shp}
P.~H. Frampton and T.~W. Kephart, \emph{{The Amaterasu Cosmic Ray as a Magnetic
  Monopole and Implications for Extensions of the Standard Model}},
  \href{https://arxiv.org/abs/2403.12322}{{\ttfamily 2403.12322}}.

\bibitem{ParticleDataGroup:2022pth}
{\scshape Particle Data Group} collaboration, R.~L. Workman et~al.,
  \emph{{Review of Particle Physics}},
  \href{https://doi.org/10.1093/ptep/ptac097}{\emph{PTEP} {\bfseries 2022}
  (2022) 083C01}.

\bibitem{Kibble:1976sj}
T.~W.~B. Kibble, \emph{{Topology of Cosmic Domains and Strings}},
  \href{https://doi.org/10.1088/0305-4470/9/8/029}{\emph{J. Phys. A} {\bfseries
  9} (1976) 1387--1398}.

\bibitem{Zurek:1985qw}
W.~H. Zurek, \emph{{Cosmological Experiments in Superfluid Helium?}},
  \href{https://doi.org/10.1038/317505a0}{\emph{Nature} {\bfseries 317} (1985)
  505--508}.

\bibitem{Murayama:2009nj}
H.~Murayama and J.~Shu, \emph{{Topological Dark Matter}},
  \href{https://doi.org/10.1016/j.physletb.2010.02.037}{\emph{Phys. Lett. B}
  {\bfseries 686} (2010) 162--165},
  [\href{https://arxiv.org/abs/0905.1720}{{\ttfamily 0905.1720}}].

\bibitem{Das:2021wei}
S.~Das and A.~Hook, \emph{{Black hole production of monopoles in the early
  universe}}, \href{https://doi.org/10.1007/JHEP12(2021)145}{\emph{JHEP}
  {\bfseries 12} (2021) 145},
  [\href{https://arxiv.org/abs/2109.00039}{{\ttfamily 2109.00039}}].

\bibitem{Kobayashi:2021des}
T.~Kobayashi, \emph{{Monopole-antimonopole pair production in primordial
  magnetic fields}},
  \href{https://doi.org/10.1103/PhysRevD.104.043501}{\emph{Phys. Rev. D}
  {\bfseries 104} (2021) 043501},
  [\href{https://arxiv.org/abs/2105.12776}{{\ttfamily 2105.12776}}].

\bibitem{Kobayashi:2022qpl}
T.~Kobayashi and D.~Perri, \emph{{Parker bound and monopole pair production
  from primordial magnetic fields}},
  \href{https://doi.org/10.1103/PhysRevD.106.063016}{\emph{Phys. Rev. D}
  {\bfseries 106} (2022) 063016},
  [\href{https://arxiv.org/abs/2207.08246}{{\ttfamily 2207.08246}}].

\bibitem{Maji:2022jzu}
R.~Maji and Q.~Shafi, \emph{{Monopoles, strings and gravitational waves in
  non-minimal inflation}},
  \href{https://doi.org/10.1088/1475-7516/2023/03/007}{\emph{JCAP} {\bfseries
  03} (2023) 007}, [\href{https://arxiv.org/abs/2208.08137}{{\ttfamily
  2208.08137}}].

\bibitem{He:2022wwy}
M.~He, K.~Kohri, K.~Mukaida and M.~Yamada, \emph{{Formation of hot spots around
  small primordial black holes}},
  \href{https://doi.org/10.1088/1475-7516/2023/01/027}{\emph{JCAP} {\bfseries
  01} (2023) 027}, [\href{https://arxiv.org/abs/2210.06238}{{\ttfamily
  2210.06238}}].

\bibitem{Lazarides:2024niy}
G.~Lazarides, R.~Maji and Q.~Shafi, \emph{{Quantum tunneling in the early
  universe: Stable magnetic monopoles from metastable cosmic strings}},
  \href{https://arxiv.org/abs/2402.03128}{{\ttfamily 2402.03128}}.

\bibitem{Epele:2007ic}
L.~N. Epele, H.~Fanchiotti, C.~A. Garcia~Canal and V.~Vento, \emph{{Monopolium:
  The Key to monopoles}},
  \href{https://doi.org/10.1140/epjc/s10052-008-0628-0}{\emph{Eur. Phys. J. C}
  {\bfseries 56} (2008) 87--95},
  [\href{https://arxiv.org/abs/hep-ph/0701133}{{\ttfamily hep-ph/0701133}}].

\bibitem{Senoguz:2015lba}
V.~N. \c{S}eno\u{g}uz and Q.~Shafi, \emph{{Primordial monopoles, proton decay,
  gravity waves and GUT inflation}},
  \href{https://doi.org/10.1016/j.physletb.2015.11.037}{\emph{Phys. Lett. B}
  {\bfseries 752} (2016) 169--174},
  [\href{https://arxiv.org/abs/1510.04442}{{\ttfamily 1510.04442}}].

\bibitem{Kephart:2017esj}
T.~W. Kephart, G.~K. Leontaris and Q.~Shafi, \emph{{Magnetic Monopoles and Free
  Fractionally Charged States at Accelerators and in Cosmic Rays}},
  \href{https://doi.org/10.1007/JHEP10(2017)176}{\emph{JHEP} {\bfseries 10}
  (2017) 176}, [\href{https://arxiv.org/abs/1707.08067}{{\ttfamily
  1707.08067}}].

\bibitem{Lazarides:2019xai}
G.~Lazarides and Q.~Shafi, \emph{{Monopoles, Strings, and Necklaces in $SO(10)$
  and $E_6$}}, \href{https://doi.org/10.1007/JHEP10(2019)193}{\emph{JHEP}
  {\bfseries 10} (2019) 193},
  [\href{https://arxiv.org/abs/1904.06880}{{\ttfamily 1904.06880}}].

\bibitem{Chakrabortty:2020otp}
J.~Chakrabortty, G.~Lazarides, R.~Maji and Q.~Shafi, \emph{{Primordial
  Monopoles and Strings, Inflation, and Gravity Waves}},
  \href{https://doi.org/10.1007/JHEP02(2021)114}{\emph{JHEP} {\bfseries 02}
  (2021) 114}, [\href{https://arxiv.org/abs/2011.01838}{{\ttfamily
  2011.01838}}].

\bibitem{Vento:2020vsq}
V.~Vento, \emph{{Primordial monopolium as dark matter}},
  \href{https://doi.org/10.1140/epjc/s10052-021-09027-6}{\emph{Eur. Phys. J. C}
  {\bfseries 81} (2021) 229},
  [\href{https://arxiv.org/abs/2011.10327}{{\ttfamily 2011.10327}}].

\bibitem{Brandenberger:2009jq}
R.~H. Brandenberger, \emph{{Alternatives to the inflationary paradigm of
  structure formation}},
  \href{https://doi.org/10.1142/S2010194511000109}{\emph{Int. J. Mod. Phys.
  Conf. Ser.} {\bfseries 01} (2011) 67--79},
  [\href{https://arxiv.org/abs/0902.4731}{{\ttfamily 0902.4731}}].

\bibitem{Rubakov:1982fp}
V.~A. Rubakov, \emph{{Adler-Bell-Jackiw Anomaly and Fermion Number Breaking in
  the Presence of a Magnetic Monopole}},
  \href{https://doi.org/10.1016/0550-3213(82)90034-7}{\emph{Nucl. Phys. B}
  {\bfseries 203} (1982) 311--348}.

\bibitem{Callan:1982ah}
C.~G. Callan, Jr., \emph{{Disappearing Dyons}},
  \href{https://doi.org/10.1103/PhysRevD.25.2141}{\emph{Phys. Rev. D}
  {\bfseries 25} (1982) 2141}.

\bibitem{Callan:1982au}
C.~G. Callan, Jr., \emph{{Dyon-Fermion Dynamics}},
  \href{https://doi.org/10.1103/PhysRevD.26.2058}{\emph{Phys. Rev. D}
  {\bfseries 26} (1982) 2058--2068}.

\bibitem{Medvedev:2017jdn}
M.~V. Medvedev and A.~Loeb, \emph{{Plasma Constraints on the Cosmological
  Abundance of Magnetic Monopoles and the Origin of Cosmic Magnetic Fields}},
  \href{https://doi.org/10.1088/1475-7516/2017/06/058}{\emph{JCAP} {\bfseries
  06} (2017) 058}, [\href{https://arxiv.org/abs/1704.05094}{{\ttfamily
  1704.05094}}].

\bibitem{Spicer:1984uui}
D.~S. Spicer and R.~N. Sudan, \emph{{Beam-return current systems in solar
  flares}}, \href{https://doi.org/10.1086/162011}{\emph{Astrophysical Journal}
  {\bfseries 280} (May, 1984) 448--456}.

\bibitem{Bai:2020spd}
Y.~Bai, J.~Berger, M.~Korwar and N.~Orlofsky, \emph{{Phenomenology of magnetic
  black holes with electroweak-symmetric coronas}},
  \href{https://doi.org/10.1007/JHEP10(2020)210}{\emph{JHEP} {\bfseries 10}
  (2020) 210}, [\href{https://arxiv.org/abs/2007.03703}{{\ttfamily
  2007.03703}}].

\bibitem{Thorne:2017mcp}
K.~S. Thorne and R.~D. Blandford, \emph{{Modern Classical Physics}}.
\newblock Princeton University Press, Princeton and Oxford.

\bibitem{Parker:1987osc}
E.~N. Parker, \emph{{Magnetic Monopole Plasma Oscillations and the Survival of
  Galactic Magnetic Fields}},
  \href{https://doi.org/10.1086/165633}{\emph{Astrophysical Journal} {\bfseries
  321} (Oct., 1987) 349}.

\bibitem{Parker:1970xv}
E.~N. Parker, \emph{{The Origin of Magnetic Fields}},
  \href{https://doi.org/10.1086/150442}{\emph{Astrophys. J.} {\bfseries 160}
  (1970) 383}.

\bibitem{Turner:1982ag}
M.~S. Turner, E.~N. Parker and T.~J. Bogdan, \emph{{Magnetic Monopoles and the
  Survival of Galactic Magnetic Fields}},
  \href{https://doi.org/10.1103/PhysRevD.26.1296}{\emph{Phys. Rev. D}
  {\bfseries 26} (1982) 1296}.

\bibitem{Weinberg:2003ur}
S.~Weinberg, \emph{{Damping of tensor modes in cosmology}},
  \href{https://doi.org/10.1103/PhysRevD.69.023503}{\emph{Phys. Rev. D}
  {\bfseries 69} (2004) 023503},
  [\href{https://arxiv.org/abs/astro-ph/0306304}{{\ttfamily
  astro-ph/0306304}}].

\bibitem{Flauger:2017ged}
R.~Flauger and S.~Weinberg, \emph{{Gravitational Waves in Cold Dark Matter}},
  \href{https://doi.org/10.1103/PhysRevD.97.123506}{\emph{Phys. Rev. D}
  {\bfseries 97} (2018) 123506},
  [\href{https://arxiv.org/abs/1801.00386}{{\ttfamily 1801.00386}}].

\bibitem{Bret:2015qia}
A.~Bret, \emph{{Collisional behaviors of astrophysical collisionless plasmas}},
  \href{https://doi.org/10.1017/S0022377815000173}{\emph{J. Plasma Phys.}
  {\bfseries 81} (2015) 455810202},
  [\href{https://arxiv.org/abs/1502.00626}{{\ttfamily 1502.00626}}].

\bibitem{2008gady.book.....B}
J.~{Binney} and S.~{Tremaine}, \emph{{Galactic Dynamics: Second Edition}}.
\newblock 2008.

\bibitem{Hansen:2004dg}
S.~H. Hansen, D.~Egli, L.~Hollenstein and C.~Salzmann, \emph{{Dark matter
  distribution function from non-extensive statistical mechanics}},
  \href{https://doi.org/10.1016/j.newast.2005.01.005}{\emph{New Astron.}
  {\bfseries 10} (2005) 379},
  [\href{https://arxiv.org/abs/astro-ph/0407111}{{\ttfamily
  astro-ph/0407111}}].

\bibitem{Ling:2009eh}
F.~S. Ling, E.~Nezri, E.~Athanassoula and R.~Teyssier, \emph{{Dark Matter
  Direct Detection Signals inferred from a Cosmological N-body Simulation with
  Baryons}}, \href{https://doi.org/10.1088/1475-7516/2010/02/012}{\emph{JCAP}
  {\bfseries 02} (2010) 012},
  [\href{https://arxiv.org/abs/0909.2028}{{\ttfamily 0909.2028}}].

\bibitem{2013SSRv..175..183L}
G.~{Livadiotis} and D.~J. {McComas}, \emph{{Understanding Kappa Distributions:
  A Toolbox for Space Science and Astrophysics}},
  \href{https://doi.org/10.1007/s11214-013-9982-9}{\emph{Space Science Reviews}
  {\bfseries 175} (June, 2013) 183--214}.

\bibitem{2014JPlPh..80..341L}
G.~{Livadiotis}, D.~J. {McComas} and {McComas}, \emph{{Electrostatic shielding
  in plasmas and the physical meaning of the Debye length}},
  \href{https://doi.org/10.1017/S0022377813001335}{\emph{Journal of Plasma
  Physics} {\bfseries 80} (June, 2014) 341--378}.

\bibitem{Burdin:2014xma}
S.~Burdin, M.~Fairbairn, P.~Mermod, D.~Milstead, J.~Pinfold, T.~Sloan et~al.,
  \emph{{Non-collider searches for stable massive particles}},
  \href{https://doi.org/10.1016/j.physrep.2015.03.004}{\emph{Phys. Rept.}
  {\bfseries 582} (2015) 1--52},
  [\href{https://arxiv.org/abs/1410.1374}{{\ttfamily 1410.1374}}].

\bibitem{Kobayashi:2023ryr}
T.~Kobayashi and D.~Perri, \emph{{Parker bounds on monopoles with arbitrary
  charge from galactic and primordial magnetic fields}},
  \href{https://doi.org/10.1103/PhysRevD.108.083005}{\emph{Phys. Rev. D}
  {\bfseries 108} (2023) 083005},
  [\href{https://arxiv.org/abs/2307.07553}{{\ttfamily 2307.07553}}].

\bibitem{Gould:2017zwi}
O.~Gould and A.~Rajantie, \emph{{Magnetic monopole mass bounds from heavy ion
  collisions and neutron stars}},
  \href{https://doi.org/10.1103/PhysRevLett.119.241601}{\emph{Phys. Rev. Lett.}
  {\bfseries 119} (2017) 241601},
  [\href{https://arxiv.org/abs/1705.07052}{{\ttfamily 1705.07052}}].

\bibitem{Hook:2017vyc}
A.~Hook and J.~Huang, \emph{{Bounding millimagnetically charged particles with
  magnetars}}, \href{https://doi.org/10.1103/PhysRevD.96.055010}{\emph{Phys.
  Rev. D} {\bfseries 96} (2017) 055010},
  [\href{https://arxiv.org/abs/1705.01107}{{\ttfamily 1705.01107}}].

\bibitem{Ghosh:2020tdu}
D.~Ghosh, A.~Thalapillil and F.~Ullah, \emph{{Astrophysical hints for magnetic
  black holes}}, \href{https://doi.org/10.1103/PhysRevD.103.023006}{\emph{Phys.
  Rev. D} {\bfseries 103} (2021) 023006},
  [\href{https://arxiv.org/abs/2009.03363}{{\ttfamily 2009.03363}}].

\bibitem{Diamond:2021scl}
M.~D. Diamond and D.~E. Kaplan, \emph{{Constraints on relic magnetic black
  holes}}, \href{https://doi.org/10.1007/JHEP03(2022)157}{\emph{JHEP}
  {\bfseries 03} (2022) 157},
  [\href{https://arxiv.org/abs/2103.01850}{{\ttfamily 2103.01850}}].

\bibitem{MACRO:2002jdv}
{\scshape MACRO} collaboration, M.~Ambrosio et~al., \emph{{Final results of
  magnetic monopole searches with the MACRO experiment}},
  \href{https://doi.org/10.1140/epjc/s2002-01046-9}{\emph{Eur. Phys. J. C}
  {\bfseries 25} (2002) 511--522},
  [\href{https://arxiv.org/abs/hep-ex/0207020}{{\ttfamily hep-ex/0207020}}].

\bibitem{Planck:2018vyg}
{\scshape Planck} collaboration, N.~Aghanim et~al., \emph{{Planck 2018 results.
  VI. Cosmological parameters}},
  \href{https://doi.org/10.1051/0004-6361/201833910}{\emph{Astron. Astrophys.}
  {\bfseries 641} (2020) A6},
  [\href{https://arxiv.org/abs/1807.06209}{{\ttfamily 1807.06209}}].

\bibitem{Sofue:2020rnl}
Y.~Sofue, \emph{{Rotation Curve of the Milky Way and the Dark Matter Density}},
  \href{https://doi.org/10.3390/galaxies8020037}{\emph{Galaxies} {\bfseries 8}
  (2020) 37}, [\href{https://arxiv.org/abs/2004.11688}{{\ttfamily
  2004.11688}}].

\bibitem{Wick:2000yc}
S.~D. Wick, T.~W. Kephart, T.~J. Weiler and P.~L. Biermann, \emph{{Signatures
  for a cosmic flux of magnetic monopoles}},
  \href{https://doi.org/10.1016/S0927-6505(02)00200-1}{\emph{Astropart. Phys.}
  {\bfseries 18} (2003) 663--687},
  [\href{https://arxiv.org/abs/astro-ph/0001233}{{\ttfamily
  astro-ph/0001233}}].

\bibitem{Perri:2023ncd}
D.~Perri, K.~Bondarenko, M.~Doro and T.~Kobayashi, \emph{{Monopole acceleration
  in intergalactic magnetic fields}},
  \href{https://arxiv.org/abs/2401.00560}{{\ttfamily 2401.00560}}.

\bibitem{PierreAuger:2016imq}
{\scshape Pierre Auger} collaboration, A.~Aab et~al., \emph{{Search for
  ultrarelativistic magnetic monopoles with the Pierre Auger Observatory}},
  \href{https://doi.org/10.1103/PhysRevD.94.082002}{\emph{Phys. Rev. D}
  {\bfseries 94} (2016) 082002},
  [\href{https://arxiv.org/abs/1609.04451}{{\ttfamily 1609.04451}}].

\bibitem{Beck:2011he}
R.~Beck, \emph{{Cosmic Magnetic Fields: Observations and Prospects}},
  \href{https://doi.org/10.1063/1.3635828}{\emph{AIP Conf. Proc.} {\bfseries
  1381} (2011) 117--136}, [\href{https://arxiv.org/abs/1104.3749}{{\ttfamily
  1104.3749}}].

\bibitem{annurev:/content/journals/10.1146/annurev-astro-091916-055221}
J.~Han, \emph{Observing interstellar and intergalactic magnetic fields},
  \href{https://doi.org/https://doi.org/10.1146/annurev-astro-091916-055221}{\emph{Annual
  Review of Astronomy and Astrophysics} {\bfseries 55} (2017) 111--157}.

\bibitem{Vachaspati:2020blt}
T.~Vachaspati, \emph{{Progress on cosmological magnetic fields}},
  \href{https://doi.org/10.1088/1361-6633/ac03a9}{\emph{Rept. Prog. Phys.}
  {\bfseries 84} (2021) 074901},
  [\href{https://arxiv.org/abs/2010.10525}{{\ttfamily 2010.10525}}].

\bibitem{Rephaeli:1982nv}
Y.~Rephaeli and M.~S. Turner, \emph{{The Magnetic Monopole Flux and the
  Survival of Intracluster Magnetic Fields}},
  \href{https://doi.org/10.1016/0370-2693(83)90897-3}{\emph{Phys. Lett. B}
  {\bfseries 121} (1983) 115--118}.

\bibitem{Adams:1993fj}
F.~C. Adams, M.~Fatuzzo, K.~Freese, G.~Tarle, R.~Watkins and M.~S. Turner,
  \emph{{Extension of the Parker bound on the flux of magnetic monopoles}},
  \href{https://doi.org/10.1103/PhysRevLett.70.2511}{\emph{Phys. Rev. Lett.}
  {\bfseries 70} (1993) 2511--2514}.

\bibitem{Lewis:1999zm}
M.~J. Lewis, K.~Freese and G.~Tarle, \emph{{Protogalactic extension of the
  Parker bound}}, \href{https://doi.org/10.1103/PhysRevD.62.025002}{\emph{Phys.
  Rev. D} {\bfseries 62} (2000) 025002},
  [\href{https://arxiv.org/abs/astro-ph/9911095}{{\ttfamily
  astro-ph/9911095}}].

\bibitem{Han:1998he}
J.~L. Han, R.~Beck and E.~M. Berkhuijsen, \emph{{New clues to the magnetic
  field structure of m31}}, {\emph{Astron. Astrophys.} {\bfseries 335} (1998)
  1117}, [\href{https://arxiv.org/abs/astro-ph/9805023}{{\ttfamily
  astro-ph/9805023}}].

\bibitem{Fletcher:2003ec}
A.~Fletcher, E.~M. Berkhuijsen, R.~Beck and A.~Shukurov, \emph{{The Magnetic
  field of M 31 from multi-wavelength radio polarization observations}},
  \href{https://doi.org/10.1051/0004-6361:20034133}{\emph{Astron. Astrophys.}
  {\bfseries 414} (2004) 53--67},
  [\href{https://arxiv.org/abs/astro-ph/0310258}{{\ttfamily
  astro-ph/0310258}}].

\bibitem{Arshakian:2008cx}
T.~G. Arshakian, R.~Beck, M.~Krause and D.~Sokoloff, \emph{{Evolution of
  magnetic fields in galaxies and future observational tests with the Square
  Kilometre Array}},
  \href{https://doi.org/10.1051/0004-6361:200810964}{\emph{Astron. Astrophys.}
  {\bfseries 494} (2009) 21},
  [\href{https://arxiv.org/abs/0810.3114}{{\ttfamily 0810.3114}}].

\bibitem{Tamm:2012hw}
A.~Tamm, E.~Tempel, P.~Tenjes, O.~Tihhonova and T.~Tuvikene, \emph{{Stellar
  mass map and dark matter distribution in M31}},
  \href{https://doi.org/10.1051/0004-6361/201220065}{\emph{Astron. Astrophys.}
  {\bfseries 546} (2012) A4},
  [\href{https://arxiv.org/abs/1208.5712}{{\ttfamily 1208.5712}}].

\bibitem{Long:2015cza}
A.~J. Long and T.~Vachaspati, \emph{{Implications of a Primordial Magnetic
  Field for Magnetic Monopoles, Axions, and Dirac Neutrinos}},
  \href{https://doi.org/10.1103/PhysRevD.91.103522}{\emph{Phys. Rev. D}
  {\bfseries 91} (2015) 103522},
  [\href{https://arxiv.org/abs/1504.03319}{{\ttfamily 1504.03319}}].

\bibitem{Sorkin:1983ns}
R.~d. Sorkin, \emph{{Kaluza-Klein Monopole}},
  \href{https://doi.org/10.1103/PhysRevLett.51.87}{\emph{Phys. Rev. Lett.}
  {\bfseries 51} (1983) 87--90}.

\bibitem{Gross:1983hb}
D.~J. Gross and M.~J. Perry, \emph{{Magnetic Monopoles in Kaluza-Klein
  Theories}}, \href{https://doi.org/10.1016/0550-3213(83)90462-5}{\emph{Nucl.
  Phys. B} {\bfseries 226} (1983) 29--48}.

\bibitem{Harlow:2022ich}
D.~Harlow, B.~Heidenreich, M.~Reece and T.~Rudelius, \emph{{Weak gravity
  conjecture}}, \href{https://doi.org/10.1103/RevModPhys.95.035003}{\emph{Rev.
  Mod. Phys.} {\bfseries 95} (2023) 035003},
  [\href{https://arxiv.org/abs/2201.08380}{{\ttfamily 2201.08380}}].

\bibitem{Schneider:2010jr}
A.~Schneider, L.~Krauss and B.~Moore, \emph{{Impact of Dark Matter Microhalos
  on Signatures for Direct and Indirect Detection}},
  \href{https://doi.org/10.1103/PhysRevD.82.063525}{\emph{Phys. Rev. D}
  {\bfseries 82} (2010) 063525},
  [\href{https://arxiv.org/abs/1004.5432}{{\ttfamily 1004.5432}}].

\bibitem{Vogelsberger:2010gd}
M.~Vogelsberger and S.~D.~M. White, \emph{{Streams and caustics: the
  fine-grained structure of LCDM haloes}},
  \href{https://doi.org/10.1111/j.1365-2966.2011.18224.x}{\emph{Mon. Not. Roy.
  Astron. Soc.} {\bfseries 413} (2011) 1419},
  [\href{https://arxiv.org/abs/1002.3162}{{\ttfamily 1002.3162}}].

\bibitem{Pati:1974yy}
J.~C. Pati and A.~Salam, \emph{{Lepton Number as the Fourth Color}},
  \href{https://doi.org/10.1103/PhysRevD.10.275}{\emph{Phys. Rev. D} {\bfseries
  10} (1974) 275--289}.

\bibitem{Hartmann:2014fya}
F.~Hartmann, W.~Kilian and K.~Schnitter, \emph{{Multiple Scales in Pati-Salam
  Unification Models}},
  \href{https://doi.org/10.1007/JHEP05(2014)064}{\emph{JHEP} {\bfseries 05}
  (2014) 064}, [\href{https://arxiv.org/abs/1401.7891}{{\ttfamily 1401.7891}}].

\bibitem{DiLuzio:2020xgc}
L.~Di~Luzio, \emph{{Pati-Salam Axion}},
  \href{https://doi.org/10.1007/JHEP07(2020)071}{\emph{JHEP} {\bfseries 07}
  (2020) 071}, [\href{https://arxiv.org/abs/2005.00012}{{\ttfamily
  2005.00012}}].

\bibitem{Dolan:2020doe}
M.~J. Dolan, T.~P. Dutka and R.~R. Volkas, \emph{{Lowering the scale of
  Pati-Salam breaking through seesaw mixing}},
  \href{https://doi.org/10.1007/JHEP05(2021)199}{\emph{JHEP} {\bfseries 05}
  (2021) 199}, [\href{https://arxiv.org/abs/2012.05976}{{\ttfamily
  2012.05976}}].

\bibitem{Cacciapaglia:2019dsq}
G.~Cacciapaglia, S.~Vatani and C.~Zhang, \emph{{Composite Higgs Meets Planck
  Scale: Partial Compositeness from Partial Unification}},
  \href{https://doi.org/10.1016/j.physletb.2021.136177}{\emph{Phys. Lett. B}
  {\bfseries 815} (2021) 136177},
  [\href{https://arxiv.org/abs/1911.05454}{{\ttfamily 1911.05454}}].

\bibitem{Cacciapaglia:2020jvj}
G.~Cacciapaglia, S.~Vatani and C.~Zhang, \emph{{The Techni-Pati-Salam Composite
  Higgs}}, \href{https://doi.org/10.1103/PhysRevD.103.055001}{\emph{Phys. Rev.
  D} {\bfseries 103} (2021) 055001},
  [\href{https://arxiv.org/abs/2005.12302}{{\ttfamily 2005.12302}}].

\bibitem{Lazarides:2021tua}
G.~Lazarides and Q.~Shafi, \emph{{Triply Charged Monopole and Magnetic
  Quarks}}, \href{https://doi.org/10.1016/j.physletb.2021.136363}{\emph{Phys.
  Lett. B} {\bfseries 818} (2021) 136363},
  [\href{https://arxiv.org/abs/2101.01412}{{\ttfamily 2101.01412}}].

\bibitem{Stojkovic:2004hz}
D.~Stojkovic and K.~Freese, \emph{{A Black hole solution to the cosmological
  monopole problem}},
  \href{https://doi.org/10.1016/j.physletb.2004.12.019}{\emph{Phys. Lett. B}
  {\bfseries 606} (2005) 251--257},
  [\href{https://arxiv.org/abs/hep-ph/0403248}{{\ttfamily hep-ph/0403248}}].

\bibitem{Zhang:2023tfv}
C.~Zhang and X.~Zhang, \emph{{Gravitational capture of magnetic monopoles by
  primordial black holes in the early universe}},
  \href{https://doi.org/10.1007/JHEP10(2023)037}{\emph{JHEP} {\bfseries 10}
  (2023) 037}, [\href{https://arxiv.org/abs/2302.07002}{{\ttfamily
  2302.07002}}].

\bibitem{Zhang:2023zmb}
C.~Zhang and X.~Zhang, \emph{{Magnetic monopole meets primordial black hole: an
  extended analysis}},
  \href{https://doi.org/10.1140/epjc/s10052-024-12383-8}{\emph{Eur. Phys. J. C}
  {\bfseries 84} (2024) 100},
  [\href{https://arxiv.org/abs/2308.07166}{{\ttfamily 2308.07166}}].

\bibitem{Wang:2023qxj}
X.-Z. Wang and C.-M. Deng, \emph{{The primordial black holes solution to the
  cosmological monopole problem}},
  \href{https://doi.org/10.1140/epjc/s10052-024-12387-4}{\emph{Eur. Phys. J. C}
  {\bfseries 84} (2024) 31},
  [\href{https://arxiv.org/abs/2401.00555}{{\ttfamily 2401.00555}}].

\bibitem{Dvali:1997sa}
G.~R. Dvali, H.~Liu and T.~Vachaspati, \emph{{Sweeping away the monopole
  problem}}, \href{https://doi.org/10.1103/PhysRevLett.80.2281}{\emph{Phys.
  Rev. Lett.} {\bfseries 80} (1998) 2281--2284},
  [\href{https://arxiv.org/abs/hep-ph/9710301}{{\ttfamily hep-ph/9710301}}].

\bibitem{Brush:2015vda}
M.~Brush, L.~Pogosian and T.~Vachaspati, \emph{{Magnetic
  monopole\textemdash{}domain wall collisions}},
  \href{https://doi.org/10.1103/PhysRevD.92.045008}{\emph{Phys. Rev. D}
  {\bfseries 92} (2015) 045008},
  [\href{https://arxiv.org/abs/1505.08170}{{\ttfamily 1505.08170}}].

\bibitem{Huang:2020bbe}
W.-C. Huang, F.~Sannino and Z.-W. Wang, \emph{{Gravitational Waves from
  Pati-Salam Dynamics}},
  \href{https://doi.org/10.1103/PhysRevD.102.095025}{\emph{Phys. Rev. D}
  {\bfseries 102} (2020) 095025},
  [\href{https://arxiv.org/abs/2004.02332}{{\ttfamily 2004.02332}}].

\bibitem{Lozanov:2023aez}
K.~D. Lozanov, M.~Sasaki and V.~Takhistov, \emph{{Universal Gravitational Wave
  Signatures of Cosmological Solitons}},
  \href{https://arxiv.org/abs/2304.06709}{{\ttfamily 2304.06709}}.

\bibitem{Lozanov:2023knf}
K.~D. Lozanov, M.~Sasaki and V.~Takhistov, \emph{{Universal gravitational waves
  from interacting and clustered solitons}},
  \href{https://doi.org/10.1016/j.physletb.2023.138392}{\emph{Phys. Lett. B}
  {\bfseries 848} (2024) 138392},
  [\href{https://arxiv.org/abs/2309.14193}{{\ttfamily 2309.14193}}].

\end{thebibliography}\endgroup
\bibliographystyle{JHEP}

\end{document}